\documentclass[aps,prd,twocolumn,showpacs,preprintnumbers,superscriptaddress,nofootinbib,floatfix]{revtex4-1}

\usepackage{epsf}
\usepackage{amsmath}
\usepackage{amssymb}
\usepackage{amsfonts}
\usepackage{dcolumn}
\usepackage{natbib}
\usepackage{bm}
\usepackage{graphicx}
\usepackage{enumerate}
\usepackage{txfonts}

\newcommand{\w}[1]{\bm{#1}}
\newcommand{\be}{\begin{equation}}
\newcommand{\ee}{\end{equation}}
\newcommand{\bea}{\begin{eqnarray}}
\newcommand{\eea}{\end{eqnarray}}

\begin{document}

\title{Gravitational waves in dynamical spacetimes with matter content in the 
Fully Constrained Formulation}

\newcommand*{\VAL}{Departamento de Astronom\'\i a y Astrof\'\i sica, 
Universidad de Valencia, C/ Dr.~Moliner 50, E-46100 Burjassot, Valencia, Spain}

\newcommand*{\MPA}{Max-Planck-Institute for Astrophysics, Garching, 
Karl-Schwarschild-Str. 1, D-85741 Garching, Germany}

\author{Isabel Cordero-Carri\'on}
\email{chabela@mpa-garching.mpg.de}
\affiliation{\MPA}
\affiliation{\VAL}

\author{Pablo Cerd\'a-Dur\'an}
\email{cerda@mpa-garching.mpg.de}
\affiliation{\MPA}
\affiliation{\VAL}

\author{Jos\'e Mar\'ia Ib\'a\~nez}
\email{Jose.M.Ibanez@uv.es}
\affiliation{\VAL}

\date{\today}

\hyphenation{pro-per-ties asso-cia-ted e-llip-tic im-po-sing taking boun-da-ry 
appro-xi-ma-tion me-thods a-xi-sym-me-tric re-co-ve-ry de-velop-ment 
Nu-meri-cal pro-blem ho-mo-ge-neity diffe-ren-ces differs stellar 
gra-vi-ta-tio-nal choosing de-noting slicing carried analy-sis following
va-riables linear-ly e-qui-li-brium second Between eigen-values proper-ly
con-figu-ra-tion regu-lar general pre-vious for-malism he-xa-de-ca-pole
spherical har-monics similar si-mu-la-tions in-fluence}

\begin{abstract}
The Fully Constrained Formulation (FCF) of General Relativity (GR) is a 
framework introduced as an alternative to the hyperbolic formulations 
traditionally used in numerical relativity. The FCF equations form a hybrid 
elliptic-hyperbolic system of equations including explicitly the constraints. 
We present an implicit-explicit numerical algorithm to solve the hyperbolic 
part, whereas the elliptic sector shares the form and properties with the well 
known Conformally Flat Condition (CFC) approximation. We show the stability and
convergence properties of the numerical scheme with numerical simulations of 
vacuum solutions. We have performed the first numerical evolutions of the 
coupled system of hydrodynamics and Einstein equations within FCF. As a proof 
of principle of the viability of the formalism, we present 2D axisymmetric 
simulations of an oscillating neutron star. In order to simplify the analysis 
we have neglected the back-reaction of the gravitational waves (GWs) into the 
dynamics, which is small ($<2\%$) for the system considered in this work. We 
use spherical coordinates grids which are well adapted for simulations of stars
and allow for extended grids that marginally reach the wave zone. We have 
extracted the GW signature and compared to the Newtonian quadrupole and 
hexadecapole formulae. Both extraction methods show agreement within the 
numerical errors and the approximations used ($\sim 30\%$).
\end{abstract}


\maketitle

%%%%%%%%%%%%%%%%%%%%%%%%%%%%%%%%%%%%%%

\section{Introduction}

Numerical relativity is a rather young branch of physics devoted to the 
numerical solution of Einstein equations for complex problems, mostly in
theoretical astrophysics, which involve the evolution of spacetime and 
eventually the matter content of a system. It was born thanks to the 
theoretical advances which led to the 3+1 split of the Einstein equations
\cite{Lichnerowicz44, CB52} popularized due to the work of \cite{ADM}. The 3+1
split defines a foliation of spacetime which allows to solve the equations as 
an initial boundary value problem for a given spatial hypersurface which is 
then evolved in time.

Soon after that theoretical breakthrough, the first hydrodynamic calculations 
of the general-relativistic collapse of a star in spherical symmetry using 
a Lagrangian code were performed \cite{May66, May67}. Multidimensional 
simulations had to wait for the development of an Eulerian formulation 
\cite{Wilson72} which could overcome the problems of non-spherical Lagrangian 
codes. Those advances led to a tremendously prolific era of general 
relativistic hydrodynamics with multiple applications to the formation of black
holes, accretion onto compact objects, binary neutron star mergers and core 
collapse supernovae (see \cite{Font08} for a recent review on the topic).

In parallel, a considerable effort was made to solve Einstein equations in 
vacuum. Only recently it has been possible to gain the sufficient understanding
of the stability properties of numerical solutions of Einstein equations to 
overcome the numerical challenge of simulating the merger of two black holes 
and estimate its gravitational wave (GW) signal \cite{Pretorius05b}. Despite 
the fact that some of the first successful simulations \cite{Pretorius05b, 
Lindblom06, Szilagyi07} used the Generalized Harmonic formulation (GH)
\cite{Friedrich85, Pretorius05a} of Einstein equations (see more about GH 
below), most of the groups in the numerical relativity community make use of a 
different numerical recipe, resulting from the combination of different 
ingredients: i) the so-called BSSN formulation \cite{Baumgarte98, Shibata95};
ii) the appropriate choice of gauge, with a slicing of the 1+log family and
some variant of the hyperbolic Gamma-driver condition for the spatial gauge 
\cite{Bona95, Alcubierre03, vanMeter06}; iii) the use of high order spatial 
methods (at least fourth-order); and iv) the use of high resolution due to the 
increasing computational power and development of adaptive mesh refinement 
(AMR) techniques. Using this recipe a number of groups \cite{Campanelli06, 
Baker06, Koppitz07, Herrmann07, Sperhake07, Bruegmann08} have provided 
waveforms of one of the most powerful sources of gravitational radiation in the
Universe, the binary black hole merger. These events are one of the main 
candidates for the first direct detection of GWs in the ground-based GW 
observatories (LIGO, Virgo, GEO600, TAMA300).

According to the success of present formulations of the Einstein equations in
different multidimensional scenarios, one could think that all the goals of
numerical relativity have been achieved. Indeed, according to \cite{Shapiro86},
the Holy Grail of numerical relativity is a numerical code to solve Einstein 
equations, that simultaneously avoids singularities, handles black holes, 
maintains high accuracy, and runs forever. Current codes satisfy all the above 
requirements (with the exception, perhaps, of the last one), thanks to both the
use of accurate numerical techniques and stable formulations of Einstein 
equations. There are, however, a number of problems that could arise when using
such an homogeneous set of numerical recipes to solve a complex 
multidimensional numerical problem.

The first set of problems could be related to the homogeneity itself. All 
working formulations of Einstein equations consist in the solution of 
hyperbolic equations, and in most of the cases make use of the 3+1 
decomposition of Einstein equations in a BSSN-like form, and use high order 
finite differences techniques to solve them numerically. This might make it 
difficult to detect fundamental theoretical problems in the set of equations to
be solved.

Some work has appeared facing these questions. The original (as far as solving 
the binary black hole problem) article of \cite{Pretorius05b} and some later 
ones \cite{Lindblom06, Szilagyi07} make use of GH formulation. It is a 4D 
covariant formulation which differs substantially from the 3+1 BSSN approach 
and allows for genuine comparisons between waveforms using different 
formulations \cite{Baker07,Hannam09}. Within the 3+1 decomposition of Einstein 
equations, some formulations have appeared \cite{Bona95, Alic11} in order to 
try to improve some of the weaknesses of the BSSN formulation, such as a better
preservation of the constraints. An alternative is the characteristic approach 
(see \cite{WinicourLR} for a review). The Cauchy-characteristic matching and 
extraction, based on a (2+1)+1 formulation, has been successfully used to 
accurately extract gravitational waves matching its evolution to interior 
Cauchy data \cite{Reisswig09, Babiuc11, Reisswig11}. However, the success in 
using this approach to simulate the whole spacetime is very limited 
\cite{Gomez98, Siebel02} due to the formation of coordinate caustics. Regarding
the numerical methods, finite differences techniques have been used in all 
works, except for \cite{Lindblom06} in which pseudo-spectral methods were used,
although no substantial differences have been found in comparison with finite 
differences codes \cite{Hannam09}. Nevertheless, an alternative to pure 
hyperbolic formulations of the Einstein equations, as the work that we present 
here, is interesting and desirable.

The second class of problems is related to the increasing level of complexity
in the astrophysical scenarios that the community wants to simulate. After the 
binary black hole problem has been solved, the numerical relativity community 
is now concentrating more on the problem of solving non-vacuum spacetimes. The 
collapse of stellar cores and the merger of neutron stars represent challenges 
by itself, beyond the numerical evolution of Einstein equations: realistic 
microphysics, accurate multidimensional neutrino transport, magnetic fields, 
small scale instabilities (e.g., magnetorotational instability) and turbulence,
non-ideal effects, elastic properties of the crust of neutron stars, 
superfluidity and superconductivity in cold neutron star interiors. Although 
full General Relativity (GR) is unavoidable in the case of the presence of 
black holes, approximations to GR in scenarios in which only neutron stars are 
present have a chance to simplify the numerical simulations to be able to 
understand the full physical complexity of those systems without the burden of 
solving the full GR equations.

The typical neutron star has a mass of about $M\sim 1.4 M_\odot$ and a radius 
of $R\sim 10-15$~km, which results in a compactness of $GM/Rc^2 \sim 0.2<1$. 
This implies that a post-Newtonian expansion of the gravitational field of an 
isolated neutron star is possible and convergent, and that the expected error in
the dynamics of the system, if Newtonian gravity is used instead of full GR, is 
about $20\%$. However, it is well known that GR does not only produces 
quantitative effects in the dynamics of the system, but also qualitatively new 
effects like frame dragging or GWs. These two classes of effects appear at 1 
and 2.5 post-Newtonian level, respectively (see e.g. \cite{Blanchet90}), which 
represent changes in the dynamics of neutron stars of about $20\%$ and $<2\%$ 
at most. However, these effects can be important due to the non-linearity of 
the equations. As an example, in the case of neutron star binaries the energy 
loss due to GWs, despite their small nominal effect, make the orbit shrink 
until the neutron stars merge. In the case of the collapse of stellar cores, 
the stellar interior models used as progenitors can be treated safely in the 
Newtonian limit ($GM/Rc^2 \sim 10^{-3}<<1$), and only as nuclear density is 
reached GR effects appear. In that case, it would be desirable to have 
numerical tools which could allow us to evolve efficiently and smoothly the 
spacetime, from the Newtonian regime of an stellar core to the mildly 
relativistic regime of a proto-neutron star or fully relativistic regime of a 
black hole. 

An approximation to GR that could fill this gap between Newtonian gravity and 
GR is the Conformally Flat Condition (CFC) approximation (\cite{Isenberg, 
WilsoM89}). The main features of the CFC approximation are: i) although it is 
not a post-Newtonian approximation, it behaves as a 1PN theory \cite{Kley94}, 
and hence, it is possible to recover the Newtonian limit correctly in the case 
of weak gravity; ii) in spherical symmetry it coincides with GR, which makes it
accurate for quasi-spherical objects like isolated neutron stars or for the 
collapse of stellar cores; iii) it only involves Poisson-like equations for the
spacetime, and therefore the numerical methods and computational costs are 
closer to Newtonian simulations than to full GR simulations; and iv) it 
neglects GWs and the energy losses related to them. The numerical solution of 
elliptic equations is more involved than hyperbolic equations.
However, the time-step in CFC is limited by the sound speed instead of the 
speed of light, as in the case of hyperbolic formulations of GR. That provides 
a considerably speed up in many scenarios that widely overcomes the extra cost 
of solving elliptic equations, making the numerical evolution considerably 
faster. The CFC approximation was originally thought to deal with the neutron 
star binary case \cite{Wilson96, Oechslin02, Faber04, Oechslin07}. In this 
case, energy losses by GWs have to be included as an extra ingredient to allow 
for the neutron stars to merge. The major success, however, has been in the 
collapse of stellar cores (\cite{Dimmelmeier01, Dimmelmeier02a, 
Dimmelmeier02b}), which lead to the computation of GW emission using physically
motivated microphysics \cite{Dimmelmeier09}, magnetic fields 
\cite{CerdaDuran08}, and neutrino transport \cite{Mueller10}. The CFC approach 
has also been successfully used to simulate the phase-transition-induced 
collapse of rotating neutron stars to hybrid quark stars \cite{Abdikamalov09} 
and the evolution of equilibrium models of rotating neutron stars \cite{Cook96,
Dimmelmeier06}. Direct comparisons of the CFC approach with full GR have shown 
that differences between both approaches, in the case of core collapse, are 
smaller than the numerical differences between the codes \cite{Shibata04, 
Ott07a, Ott07b}. This fact is understandable since the next post-Newtonian 
corrections to CFC were found to have an impact on the non-linear dynamics of 
less than $1\%$ \cite{CerdaDuran05}. In the case of neutron star mergers we are
not aware of a direct comparison between CFC and full GR.

A new formulation of Einstein equations which could address both classes of
problems mentioned above, and share some properties with the CFC approximation,
is the Fully Constrained Formulation (FCF) \cite{Bonazzola04}. This formulation
is based on the 3+1 split of Einstein equations but different from all other 
formulations of Einstein equations that are purely hyperbolic, the FCF 
maximizes the number of elliptic equations by solving the constraint equations 
at each time-step and choosing an appropriate gauge condition. As a 
consequence, the hyperbolic part of FCF only contains two degrees of freedom, 
which correspond, far from the matter sources, to the GW content of the system.
Therefore, FCF is fundamentally different from fully hyperbolic formulations of
GR and can be used as another check of the consistency of the numerical 
solutions of Einstein equations. There are other formulations which incorporate
the constraints into the evolution system (see \cite{JarValGou08}, section 
5.2.2., for most relevant examples); however, no numerical simulations have 
been performed yet with most of them (e.g., \cite{AnMon}), or simulations are
restricted to axisymmetric spacetimes (e.g., \cite{Rinne}). Moreover, FCF is a 
natural generalization of the CFC approximation; this fact makes possible a 
natural extension of all the numerical codes which use this approximation, in 
order to have a proper treatment of the gravitational radiation of the system 
without too much effort. It also creates a bridge between weak gravity systems,
which are well described within the CFC approximation, and the strong gravity 
limit.

In practice, to extend a CFC code to FCF one has to add additional hyperbolic 
equations to the existing CFC elliptic equations (and also some extra sources 
in these elliptic equations). This evolution system, written as a first-order 
one, is a hyperbolic system \cite{CC08b} and includes the whole hyperbolic 
sector of the metric of spacetime in this formulation. In particular, the 
explicit values of the eigenvalues and eigenvectors of the hyperbolic metric 
system allow to guarantee the expected physical behavior on trapping horizons 
\cite{CC08a}. The remaining metric variables form the elliptic sector, which is
similar to the group of elliptic equations in the CFC approximation, with extra
sources. Recent works \cite{CC09} overcome some pathological problems related 
with non-local uniqueness in the elliptic equations in CFC, and also in FCF. 
The equations were rewritten in such a way that these problems were solved, and
the new scheme has been used successfully in some applications \cite{Mueller10,
BucZan, BausThesis}.

The analysis of the numerical evolution of the hyperbolic metric system is the 
main objective of this work, including aspects like numerical stability of the 
system in long-term simulations, evolution of equilibrium configurations, or 
the influence of the elliptic equations in the system. We perform all the 
numerical simulations using the CoCoNuT code \cite{Dimmelmeier02a, 
Dimmelmeier02b, Dimmelmeier05}, which was originally designed to evolve the 
hydrodynamics equations in the dynamical spacetime of the CFC approximation. 
The code uses spherical coordinates for the evolution of both matter and 
spacetime; this is very convenient in the present work, since it allows us to 
place the outer boundary sufficiently far from the star, in order to perform an
accurate GW extraction \cite{CC10}.

In this paper we present the first accurate extraction of the gravitational 
wave signature coming from the evolution of rotating oscillating neutron stars 
within FCF. As a first step towards a full evolution of the coupled system of 
elliptic and hyperbolic equations of the FCF, we have neglected the 
back-reaction of the GWs onto the dynamics of the system, which is a justified 
approximation in the case of isolated neutron stars and the collapse of stellar
cores.

The article is organized as follows. In Sec. II we review the FCF and detail 
the formulation used in the evolution of spacetime. In Sec. III we describe the
numerical methods used in the evolution of the different systems of equations 
(hydrodynamics, elliptic and hyperbolic metric equations). In Sec. IV we test 
our numerical implementation through the evolution of a vacuum spacetime with 
analytical solution. In Sec. V we perform simulations of equilibrium 
configurations of rotating neutron star and extract GWs from perturbed 
oscillating models. Conclusions are drawn in Sec. V. Throughout the paper we 
use the signature $(-,+,+,+)$ for the spacetime metric, and units in which 
$c=G=M_\odot=1$. Greek indices run from 0 to 3, whereas Latin ones from 1 to 3 
only.

%%%%%%%%%%%%%%%%%%%%%%%%%%%%%%%%%%%%%%

\section{Spacetime evolution}

\subsection{Fully Constrained Formalism}
Given an asymptotically flat spacetime $({\cal M}, g_{\mu\nu})$ we consider a 
$3 + 1$ splitting by spacelike hypersurfaces $\Sigma_t$, denoting timelike unit
normals to $\Sigma_t$ by $n^\mu$. The spacetime on each spacelike hypersurface 
$\Sigma_t$ is described by the pair $(\gamma_{ij}, K^{ij})$, where 
$\gamma_{\mu\nu} = g_{\mu\nu} + n_\mu n_\nu$ is the Riemannian metric induced 
on $\Sigma_t$. We choose the convention 
$K_{\mu\nu} = - \frac{1}{2}{\cal L}_{\w n}\gamma_{\mu\nu}$ for the extrinsic 
curvature. With the lapse function $N$ and the shift vector $\beta^i$, the 
Lorentzian metric $g_{\mu\nu}$ can be expressed in coordinates $(x^\mu)$ as
\be
  g_{\mu\nu} \, dx^\mu \, dx^\nu =
  - N^2 \, dt^2 + \gamma_{ij} (dx^i + \beta^i \, dt)(dx^j + \beta^j \, dt).
\ee
As in \cite{Bonazzola04} we introduce a time independent flat metric $f_{ij}$, 
which satisfies ${\cal L}_{\w{t}}f_{ij} = \partial_t f_{ij} = 0$ and coincides 
with $\gamma_{ij}$ at spatial infinity. We define $\gamma := \det \gamma_{ij}$ 
and $f := \det f_{ij}$. We introduce the following conformal decomposition of 
the  spatial metric $\gamma_{ij}$:
\be
  \gamma_{ij} = \psi^4 \tilde{\gamma}_{ij}, \qquad \psi = (\gamma / f)^{1/12}.
\ee
The deviation of the conformal metric from the flat fiducial one is denoted by 
$h^{ij}$,
\be
  h^{ij} := \tilde{\gamma}^{ij} - f^{ij}.
\ee
Once the $3+1$ conformal decomposition is performed, a choice of gauge is 
needed in order to properly reformulate Einstein equations. The prescriptions 
in~\cite{Bonazzola04} are maximal slicing,
\bea
  K = 0,
\eea
and the so-called generalized Dirac gauge,
\be
 {\cal D}_i \tilde{\gamma}^{ij} = {\cal D}_i h^{ij} = 0,
\ee
where $K=\gamma^{ij}K_{ij}$ denotes the trace of the extrinsic curvature and 
${\cal D}_k$ stands for the Levi--Civita connection associated with the flat 
metric $f_{ij}$. More details can be found in~\cite{Bonazzola04}. Einstein 
equations then become a coupled elliptic-hyperbolic system: the elliptic sector
acts on the variables $\psi$, $N$, and $\beta^i$, and the hyperbolic sector 
acts on $h^{ij}$. More details of the analysis carried out for both elliptic 
and hyperbolic systems can be found in~\cite{CC08b, CC09}.

We introduce the conformal decomposition
\be
	\hat{A}^{ij} := \psi^{10} K^{ij},
\ee
and its decomposition in longitudinal and transverse-traceless parts
\be
	\hat{A}^{ij} = (LX)^{ij} + \hat{A}^{ij}_{\rm TT},
\label{e:dec_hatA}
\ee
where
\be
 (LX)^{ij}:= {\cal D}^i X^j + {\cal D}^j X^i - \frac{2}{3} f^{ij} {\cal D}_k X^k
\ee
and ${\cal D}_i \hat{A}^{ij}_{\rm TT} = 0$. These decompositions are motivated 
by the local uniqueness properties of elliptic equations shown in~\cite{CC09}. 
We define $w^{ij}_k := {\cal D}_k \tilde{\gamma}^{ij}$. The hyperbolic system 
for $h^{ij}$ can be written as a first-order evolution system for the tensors 
$\left( h^{ij}, \hat{A}^{ij}, w^{ij}_k \right)$,
\begin{widetext}
\bea
	\frac{\partial h^{ij}}{\partial t} & = & 2 N \psi^{-6} \hat{A}^{ij} + 
\beta^k w^{ij}_k - \tilde{\gamma}^{ik} {\cal D}_k \beta^j - 
\tilde{\gamma}^{kj} {\cal D}_k \beta^i + 
\frac{2}{3} \tilde{\gamma}^{ij} {\cal D}_k \beta^k, \label{e:evol_h}\\
	\frac{\partial \hat{A}^{ij}}{\partial t} & = & {\cal D}_k \left( 
\frac{N \psi^2}{2} \tilde{\gamma}^{kl} w^{ij}_l + \beta^k \hat{A}^{ij} \right) -
\hat{A}^{kj} {\cal D}_k \beta^i - \hat{A}^{ik} {\cal D}_k \beta^j +
\frac{2}{3} \hat{A}^{ij} {\cal D}_k \beta^k + 
2 N \psi^{-6} \tilde{\gamma}_{kl} \hat{A}^{ik} \hat{A}^{jl} \nonumber \\
	& & - 8 \pi N \psi^6 \left( \psi^4 S^{ij} - 
\frac{ S \tilde{\gamma}^{ij}}{3} \right) + N \left( \psi^2 \tilde{R}_*^{ij} + 
8 \tilde{\gamma}^{ik} \tilde{\gamma}^{jl} {\cal D}_k \psi {\cal D}_l \psi \right) + 
4 \psi \left( \tilde{\gamma}^{ik} \tilde{\gamma}^{jl} {\cal D}_k \psi {\cal D}_l N +
\tilde{\gamma}^{ik} \tilde{\gamma}^{jl} {\cal D}_k N {\cal D}_l \psi \right)
\nonumber \\
	& & - \frac{1}{3} \left[ N \left( \psi^2 \tilde{R} + 
8 \tilde{\gamma}^{kl} {\cal D}_k \psi {\cal D}_l \psi \right) +
8 \psi \tilde{\gamma}^{kl} {\cal D}_k \psi {\cal D}_l N \right] \tilde{\gamma}^{ij} 
\nonumber \\
	& & - \frac{1}{2} \left( \tilde{\gamma}^{ik} w^{lj}_k +
\tilde{\gamma}^{kj} w^{il}_k \right) {\cal D}_l (N \psi^2) -
\tilde{\gamma}^{ik} \tilde{\gamma}^{jl} {\cal D}_k {\cal D}_l (N \psi^2) +
\frac{1}{3} \tilde{\gamma}^{ij} \tilde{\gamma}^{kl} {\cal D}_k {\cal D}_l (N \psi^2), 
\label{e:evol_hatA} \\
\frac{\partial w^{ij}_k}{\partial t} &=&
{\cal D}_k \left ( 2 N \psi^{-6} \hat{A}^{ij} + \beta^l w^{ij}_l 
- \tilde{\gamma}^{il} {\cal D}_l \beta^j - \tilde{\gamma}^{lj} {\cal D}_l \beta^i + 
\frac{2}{3} \tilde{\gamma}^{ij} {\cal D}_l \beta^l \right )\,\, \label{e:evol_w}
\eea
where
\bea
	\tilde{R} &=&
\frac{1}{4} \tilde{\gamma}^{kl} {\cal D}_k h^{mn} {\cal D}_l \tilde{\gamma}_{mn} - 
\frac{1}{2} \tilde{\gamma}^{kl} {\cal D}_k h^{mn} {\cal D}_n \tilde{\gamma}_{ml}, \\
	\tilde{R}_*^{ij} &=& \frac{1}{2} \left[ - w_l^{ik} w_k^{jl} - 
\tilde{\gamma}_{kl} \tilde{\gamma}^{mn} w_m^{ik} w_n^{jl} +
\tilde{\gamma}_{nl} w_k^{mn} \left( \tilde{\gamma}^{ik} w_m^{jl} +
\tilde{\gamma}^{jk} w_m^{il} \right) \right] +
\frac{1}{4} \tilde{\gamma}^{ik} \tilde{\gamma}^{jl} w_k^{mn}
{\cal D}_l \tilde{\gamma}_{mn},
\eea
$S_{ij}:= T_{\mu \nu} \gamma^\mu_i \gamma^\nu_j$ is the stress tensor and 
$S:=\gamma^{ij}S_{ij}$ is its trace, $T_{\mu \nu}$ being the energy-momentum 
tensor, measured by the observer of 4-velocity $n^\mu$ (Eulerian observer). 
Moreover, the system obeys the constraint of the Dirac gauge, $w^{ij}_i=0$, and
for the determinant of the conformal metric, we obtain $\tilde{\gamma}=f$. 

The elliptic part of the FCF equations can be rewritten as
\bea
	\tilde{\gamma}^{kl} {\cal D}_k {\cal D}_l \psi &=& - 2 \pi \psi^{-1} E^* - 
\frac{\tilde{\gamma}_{il} \tilde{\gamma}_{jm} \hat{A}^{lm} \hat{A}^{ij}}{8 \psi^{7}} +
\frac{\psi \tilde{R}}{8}, \label{e:fcf_psi} \\
	\tilde{\gamma}^{kl} {\cal D}_k {\cal D}_l (N \psi) &=& \left[ 
2 \pi \psi^{-2} (E^* + 2 S^*) + \left( 
\frac{7 \tilde{\gamma}_{il} \tilde{\gamma}_{jm} \hat{A}^{lm} \hat{A}^{ij}}{8 \psi^{8}} +
\frac{\tilde{R}}{8} \right) \right] (N \psi), \label{e:fcf_npsi} \\
	\tilde{\gamma}^{kl} {\cal D}_k {\cal D}_l \beta^i \! + 
\frac{1}{3} \tilde{\gamma}^{ik} {\cal D}_k {\cal D}_l \beta^l &=&
16 \pi N \psi^{-6} \tilde{\gamma}^{ij}(S^*)_j +
\hat{A}^{ij} {\cal D}_j \left( 2 N \psi^{-6} \right) - 
2 N \psi^{-6} \Delta^i_{kl} \hat{A}^{kl}, \label{e:fcf_beta}
\eea
\end{widetext}
where $E:=T_{\mu \nu} n^\mu n^\nu$ and $S_i:=-\gamma^\mu_i T_{\mu\nu}n^\nu$ 
are, respectively, the energy density and the momentum density measured by the 
observer of 4-velocity $n^\mu$, $E^*:=\psi^6 E$, $S^*:=\psi^6 S$, 
$(S^*)_i:=\psi^6 S_i$, and
\be
	\Delta^k_{ij} = \frac{1}{2}\tilde{\gamma}^{kl} \left(
{\cal D}_i \tilde{\gamma}_{lj} + {\cal D}_j \tilde{\gamma}_{il} 
- {\cal D}_l \tilde{\gamma}_{ij} \right). 
\ee

The decomposition introduced in Eq.~(\ref{e:dec_hatA}) leads to an extra 
elliptic equation for the vector $X^i$,
\bea
	&&{\cal D}_j {\cal D}^j X^i + \frac{1}{3} {\cal D}^i {\cal D}_k X^k + 
\tilde{\gamma}^{im} \left( {\cal D}_k \tilde{\gamma}_{ml} - 
\frac{{\cal D}_m \tilde{\gamma}_{kl}}{2} \right) (LX)^{kl} \nonumber \\
  && \qquad = 8 \pi \tilde{\gamma}^{ij} (S^*)_j - \tilde{\gamma}^{im} \left( 
{\cal D}_k \tilde{\gamma}_{ml} - \frac{{\cal D}_m \tilde{\gamma}_{kl}}{2} 
\right) \hat{A}_{\mathrm{TT}}^{kl},\label{e:fcf_x}
\eea
and the evolution Eq.~(\ref{e:evol_hatA}) can be viewed as an evolution 
equation for the tensor $\hat{A}^{ij}_{\rm TT}$.

More details about the derivation of all the equations can be found 
in~\cite{CC09}. All these equations are to be solved coupled with the 
hydrodynamic equations for the evolution of matter which can be derived from
the Bianchi identities and the continuity equation,
\be
T^{\mu\nu}_{;\mu} =0 \qquad J^{\mu}_{;\mu} = 0.
\label{e:hydro}
\ee
Explicit expressions for the hydrodynamics equations for the case of a perfect 
fluid in the form that it is used in the present work can be found in 
\cite{Dimmelmeier05}.

\subsection{Passive FCF}

\begin{table}
\begin{tabular}{cccc}
	\hline & hydrodynamics & elliptic metric & hyperbolic \\ 
	Approach & equations & sector & metric sector \\ \hline
        CFC & \cite{Dimmelmeier02a} & (\ref{e:cfc_psi})-(\ref{e:cfc_x}) & no \\
        Passive FCF & \cite{Dimmelmeier02a} & (\ref{e:cfc_psi})-(\ref{e:cfc_x}) & (\ref{e:evol_h})-(\ref{e:evol_w}) \\
        FCF & \cite{Ibanez2000} & (\ref{e:fcf_psi})-(\ref{e:fcf_beta}), (\ref{e:fcf_x}) & (\ref{e:evol_h})-(\ref{e:evol_w}) \\ \hline
        Teukolsky waves & no (vacuum) & fixed Minkowsky & (\ref{e:evol_h})-(\ref{e:evol_w}) \\
        Equilibrium NS & fixed, \cite{lapming06} & fixed, \cite{lapming06} & (\ref{e:evol_h})-(\ref{e:evol_w})\\ 
        Oscillating NS & \cite{Dimmelmeier02a} & (\ref{e:cfc_psi})-(\ref{e:cfc_x}) & (\ref{e:evol_h})-(\ref{e:evol_w}) \\ \hline
\end{tabular}
\caption{Guide to the approaches to GR used in this work: first three rows
represent the approaches discussed in the theoretical part of the present work 
and the last three rows the approaches used in the numerical part. For each 
approach we provide the equations that we use for each of the three sectors 
(hydrodynamics, elliptic and hyperbolic metric sectors), a suitable reference 
where the equations can be found or a comment.}
\label{tab:approx}
\end{table}

An interesting property of the fully constrained formalism is that if 
$h^{ij}=0$, the resulting 3-metric $\gamma_{ij}$ is conformally flat. This 
condition corresponds to the well know conformally flat condition (CFC) 
approximation~\cite{Isenberg, WilsoM89} of Einstein equations. The CFC 
approximation has been proved to provide accurate evolutions of spacetimes 
including single neutron stars and core collapse 
supernovae~\cite{Shibata04, Ott07a, Ott07b}. In these scenarios the 
back-reaction of the GWs on the dynamics of the system is so small that 
$h^{ij}$ can be approximated to be zero. The main drawback of the CFC 
approximation is that the GW content is removed from the system, and the 
computation of the GW emission has to be performed approximately by means of 
the quadrupole formula. 

Since our aim is to deal with this kind of astrophysical scenarios, neutron 
stars and core collapse supernovae, in which the GWs are not important for the 
dynamics, when solving the complete FCF system, the back-reaction of the 
$h^{ij}$ tensor onto the hydrodynamics and elliptic part of the metric 
equations can be neglected. Therefore, we impose $h^{ij}=0$ in 
Eqs.~(\ref{e:fcf_psi})--(\ref{e:fcf_x}). The resulting system of elliptic 
equations
\bea
	\Delta\psi &=& -2\pi\psi^{-1} E^* 
- \frac{f_{il} f_{jm} \hat{A}^{lm} \hat{A}^{ij}}{8 \psi^{7}},
\label{e:cfc_psi} \\
	\Delta (N \psi) &=& 2\pi N  \psi^{-1}(E^* + 2 S^*) \nonumber \\
	&+& N \psi^{-7} \frac{7 f_{il} f_{jm} \hat{A}^{lm} \hat{A}^{ij}}{8},
\label{e:cfc_npsi} \\
	\Delta  \beta^i \! &+& \frac{1}{3} f^{ij} {\cal D}_j {\cal D}_k \beta^k =
{\cal D}_j \left(2 N \psi^{-6} \hat{A}^{ij} \right),
\label{e:cfc_beta} \\
	\Delta X^i &+& \frac{1}{3}f^{ij}{\cal D}_j{\cal D}_k X^k = 8\pi{f}^{ij}(S^*)_j
\label{e:cfc_x}
\eea
is identical to the CFC equations in the form described in~\cite{CC09}.

In the present work we solve the coupled evolution of the hyperbolic system for
$h^{ij}$ given by Eqs.~(\ref{e:evol_h})--(\ref{e:evol_w}), the elliptic 
approximated system for $N$, $\psi$ and $\beta^i$, given by
Eqs.~(\ref{e:cfc_psi})--(\ref{e:cfc_beta}) and the hydrodynamics equations. We 
call the new system passive FCF, in the sense that we neglect the back-reaction
of the GWs onto the dynamics of the system. Contrary to the CFC approximation, 
this approach does not neglect the GWs itself. Therefore, it allows one to 
compute the GW emission of the system directly from the spacetime evolution. 
Upper three rows of table~\ref{tab:approx} summarize the approximations used 
in the case of CFC, passive FCF and FCF.

%%%%%%%%%%%%%%%%%%%%%%%%%%%%%%%%%%%%%%

\section{Numerical methods}

We perform all the simulations of this work using the numerical code
CoCoNuT~\cite{Dimmelmeier02a, Dimmelmeier05, coconut}. We have extended this 
code, which solves the coupled evolution of the hydrodynamics equations with 
spacetime evolution in the CFC approximation, to add the new degrees of freedom
necessary for the FCF in the passive FCF approximation. In the following, we 
briefly describe the numerical methods used in the code to solve the 
hydrodynamics equations and the elliptic part of the FCF formalism. These 
methods and equations are identical to those described 
in~\cite{Dimmelmeier05, CC09}. We also describe the numerical techniques 
applied to solve the evolution of the $h^{ij}$ tensor, which is necessary to 
extend the CFC approximation to passive FCF. In all cases we consider spherical 
coordinates and axisymmetry. In order to simplify the notation, we will refer 
to the three sets of variables as hydrodynamics variables, 
${\bf U}:=(D, S_i, \tau)$ (see definitions below), elliptic-spacetime 
variables or CFC variables, ${\bf V}:=(N, \psi, \beta^i,X^i)$, and hyperbolic 
spacetime variables, ${\bf W}:=(h^{ij},\hat {A}^{ij}, w^{ij}_k)$.

\subsection{Hydrodynamics equations}
The system of Eqs.~(\ref{e:hydro}) can be cast into a system of conservation 
laws~\cite{Ibanez2000} as
\be
	\partial_t{\bf U} + \partial_i{\bf F}^i({\bf U},{\bf V})= Q({\bf U},{\bf V}).
\ee
${\bf U}:=(D, S_i, \tau)$ is the conserved variables vector, 
$D \equiv - J^\mu n_\mu$ and $\tau \equiv E - D$.

Since we are neglecting the back-reaction of the GWs onto the dynamics of the 
fluid, there is no dependence on the hyperbolic-spacetime variables ${\bf W}$ 
in the previous set of equations. We use Godunov-type schemes, which are 
suitable for solving equations written in conservative form. These schemes 
allow for a numerical evolution of the system with high accuracy in 
conservation of mass, momentum and energy, and the correct behavior at 
discontinuities, e.g. shock waves at the surface of neutron stars (see 
e.g.~\cite{Font08}). We use the Marquina flux formula~\cite{Donat1998} combined
with a second-order linear reconstruction with monotonized central slope 
limiter~\cite{Toro99}. The time update of the matter quantities relies on the 
method of lines in combination with a second-order accurate explicit 
Runge-Kutta scheme. The time step is restricted by the Courant-Friedrich-Lewi 
(CFL) condition~\cite{CFL1928}. This combination provides second-order 
convergence in a number of tests including the evolution of oscillating neutron
stars~\cite{Dimmelmeier05}, which degrades to first-order in the presence of 
discontinuities. We use spherical coordinates $(r, \theta, \varphi)$ in 
axisymmetry. The angular grid is equally spaced in $\theta$, but the radial 
grid can be non-equidistant.

\subsection{Elliptic spacetime equations}
Once the values of the hydrodynamic variables, $\bf U$, have been updated, the 
CFC metric, $\bf V$, can be updated by solving the elliptic part of the 
spacetime evolution equations. It consists in a system of Poisson-like elliptic
equations, Eqs.~(\ref{e:cfc_psi})--(\ref{e:cfc_beta}), which can be written as
\be
	\Delta {\bf V} = f ({\bf U}, {\bf V}).
\ee
This system of equations can be solved hierarchically, following the procedure 
described in~\cite{CC09}. 

We compute the numerical solution using spectral methods. The sources of the 
equations are interpolated (parabolic interpolation) from the finite difference
grid to the spectral one, where the elliptic equations are solved using the 
LORENE library for spectral methods~\cite{Lorene}. The spectral solution of the
equations is evaluated at the finite difference grid in order to update the 
metric fields $N$, $\psi$ and $\beta^i$, which are needed for the recovery of 
the primitive hydrodynamic variables (e.g., density and velocity) and the 
evolution of $h^{ij}$ (see next subsection). The spectral grid consists of 
several radial domains in spherical coordinates. Further details can be found 
in~\cite{Dimmelmeier05}.

Since the system of equations is elliptic, the Courant condition does not 
restrict the time step. Although the metric could be computed after every time 
step, in some scenarios the typical time scale of variation of the CFC 
variables is much longer than that of the hydrodynamic ones, and it is 
justified to compute ${\bf V}$ not at every hydrodynamics time step. In the 
simulations of neutron star oscillations presented in this paper we compute the
CFC part of the metric every 10th hydrodynamical time steps and use a parabolic
extrapolation between consecutive metric computations. This method has shown to
provide sufficient accuracy for this scenario~\cite{Dimmelmeier02a}.

\subsection{Hyperbolic spacetime equations}
\label{sect:hypereq}
Once we have updated the hydrodynamic variables, $\bf U$, and the CFC 
variables, $\bf V$, we solve the hyperbolic part of the spacetime evolution, 
Eqs.~(\ref{e:evol_h})--(\ref{e:evol_w}). This part contains the gravitational 
wave information of the system. It consists of evolution equations for the 
variables $\bf W$ of the form
\be
	\partial_t {\bf W} = g ({\bf W}, {\bf V}, {\bf U}).
\ee

We solve the system following a two step approach. In the first step we update 
$h^{ij}$ and $\hat{A}^{ij}$ to the next hypersurface, 
$\Sigma_{t^{n+1}}\equiv\Sigma_{t^n+\Delta t}$, denoted by an upperindex 
$(n+1)$, using only information of the previous hypersurface, $\Sigma_{t^n}$, 
denoted by $(n)$. It is therefore an explicit algorithm of the form
\bea
	\partial_t h^{ij} &=& 
S_h (h^{ij (n)}, \hat{A}^{ij (n)}, w^{ij (n)}_k, {\bf V}^{(n)}), \\
	\partial_t \hat{A}^{ij} &=& S_{\hat{A}} 
(h^{ij (n)}, \hat{A}^{ij (n)}, w^{ij (n)}_k, {\bf U}^{(n)}, {\bf V}^{(n)}),
\eea 
which can be integrated using explicit Runge-Kutta schemes.

In the second step we update $w^{ij}_k$ using an implicit-explicit approach. We
compute the sources using the values $(h^{ij (n)}, \hat{A}^{ij (n)})$ and the 
updated values of $(h^{ij (n+1)}, \hat{A}^{ij (n+1)})$ computed in the first 
step. The sources of Eq.~(\ref{e:evol_w}) can be splitted into two terms of the
form
\be
\partial_t w^{ij}_k = S_{w1} (h^{ij}, \hat{A}^{ij}, {\bf V}) 
+ S_{w2} (w^{ij}_k, {\bf V}),
\label{e:evol_w_sources}
\ee
where
\bea
S_{w1} &=& {\cal D}_k \left ( 2 N \psi^{-6} \hat{A}^{ij} 
- \tilde{\gamma}^{il} {\cal D}_l \beta^j - \tilde{\gamma}^{lj} {\cal D}_l \beta^i + 
\frac{2}{3} \tilde{\gamma}^{ij} {\cal D}_l \beta^l
\right ), \nonumber \\
S_{w2} &=& {\cal D}_k \left ( \beta^l w^{ij}_l \right ).
\eea

The first term, $S_{w1}$, does not depend explicitly on the evolved variables 
$w^{ij}_k$. The second term, $S_{w2}$, depends linearly on $w^{ij}_k$ and does 
not depend explicitly on the variables $(h^{ij}, \hat{A}^{ij})$. This property 
allows us to design a numerical algorithm to evolve $w^{ij}_k$ from 
$\Sigma_{t^{n}}$ to $\Sigma_{t^{n+1}}$, using the values of $h^{ij}$ and 
$\hat{A}^{ij}$ at $\Sigma_{t^{n+1}}$ and all other variables at 
$\Sigma_{t^{n}}$, i.e.,
\be
\partial_t w^{ij}_k = S_{w1} (h^{ij(n+1)}, \hat{A}^{ij(n+1)}, {\bf V}^{(n)}) 
+ S_{w2} (w^{ij(n)}_k, {\bf V}^{(n)}).
\ee
This scheme provides a numerically stable evolution, due to the (partially)
implicit dependence on $S_{w1}$ and explicit on $S_{w2}$. We evolve the system 
with the same Runge-Kutta schemes as in the first step, but with the 
corresponding partially implicit evaluation of the $S_{w1}$ source term 
(see Appendix~\ref{a:rk} for more details). However, this evaluation reduces 
the theoretical order of the scheme, which is observed in numerical simulations
in those scenarios where the terms 
$S_{w1} (h^{ij(n+1)}, \hat{A}^{ij(n+1)}, {\bf V}^{(n)})$ and 
$S_{w1} (h^{ij(n)}, \hat{A}^{ij(n)}, {\bf V}^{(n)})$ differ significantly. In 
practice, the reduction of the order of the method can be small as long as the 
leading term of the sources for the evolution of $w^{ij}_k$ is the one 
containing the $\hat{A}^{ij}$ tensor, as we have obtained in the evolution of 
Teukolsky waves with a method based on a fourth-order Runge-Kutta scheme (see 
Sec.~\ref{sec:teu}). In other cases the order of convergence of the method can 
reduce up to second-order within the same scheme, as we have obtained when the 
tensor $h^{ij}$ reaches stationary values in the evolution of equilibrium 
configurations of rotating neutron stars (see Sec.~\ref{sec:eq_neu_stars}). In 
a general scenario, an implicit-explicit (IMEX) Runge-Kutta 
scheme~\cite{AsRuuSpi, PaRu} could be used to prevent the reduction of the 
order of convergence, although this is beyond the scope of this paper. The use 
of implicit terms for the second step of the time integration is crucial in 
order to provide stability. We have checked that when a purely explicit 
approach is used for $w^{ij}_k$, the numerical method becomes unstable 
\cite{CCThesis}. The method becomes also unstable, when we compute $w^{ij}_k$ 
directly as spatial derivatives of $h^{ij}$. 

To solve Eqs.~(\ref{e:evol_w_sources}) we use a fourth-order explicit 
TVD Runge-Kutta scheme~\cite{SpiRuu}, together with the partially implicit 
treatment mentioned above. We use a fourth-order cell-centered Lagrange 
interpolation polynomials \cite{Abramowitz} to compute spatial derivatives, 
even for non-equidistan grids, and a forth-order Kreiss-Oliger dissipative 
term~\cite{kreiss73} to avoid the development of high frequency numerical 
noise. We impose an outgoing radiation Sommerfeld~\cite{Somm} condition to the 
linear part of the wave at the outer boundary, to prevent reflections from the 
boundary into the numerical domain. Unless other purely hyperbolic formulations
of Einstein equations, the boundary conditions for the hyperbolic part of FCF 
do not influence the preservation of the constraints, since they are solved 
separately.

The time step is determined by the Courant condition for the speed of light, 
$c$. This time step condition is more restrictive than that of the 
hydrodynamics because the fluid eigenvalues are bounded by $c$. The time step 
for ${\bf W}$ is chosen to be an integer fraction of the hydrodynamic time step, 
such that after each hydrodynamic time step ${\bf U}$ and ${\bf W}$ are 
synchronized. 

%%%%%%%%%%%%%%%%%%%%%%%%%%%%%%%%%%%%%%

\section{Teukolsky waves}
\label{sec:teu}

The first test is the evolution of Teukolsky waves~\cite{Teukolsky} which are 
solution of the linearized Einstein equations in a vacuum spacetime. We choose 
as initial data a combination of ingoing and outgoing even parity axisymmetric 
Teukolsky waves with amplitude $10^{-5}$. It provides regular initial data at 
$r=0$ which satisfies the Dirac gauge and is traceless (which is the linear 
approximation of unit determinant corresponding to orthonormal spherical 
coordinates for the conformal spatial metric $\tilde{\gamma}^{ij}$). We keep 
the background flat, i.e., $N=\psi=1$ and $\beta^i=0$. We assume symmetry with 
respect to the equatorial plane. The radial interval $[0,10]$ and the angular 
one $[0,\pi/2]$ are discretized by $n_r$ and $n_\theta$ equally spaced grid 
points, respectively. Table~\ref{tab:approx} summarizes the approximations made
in this test.

\begin{figure}
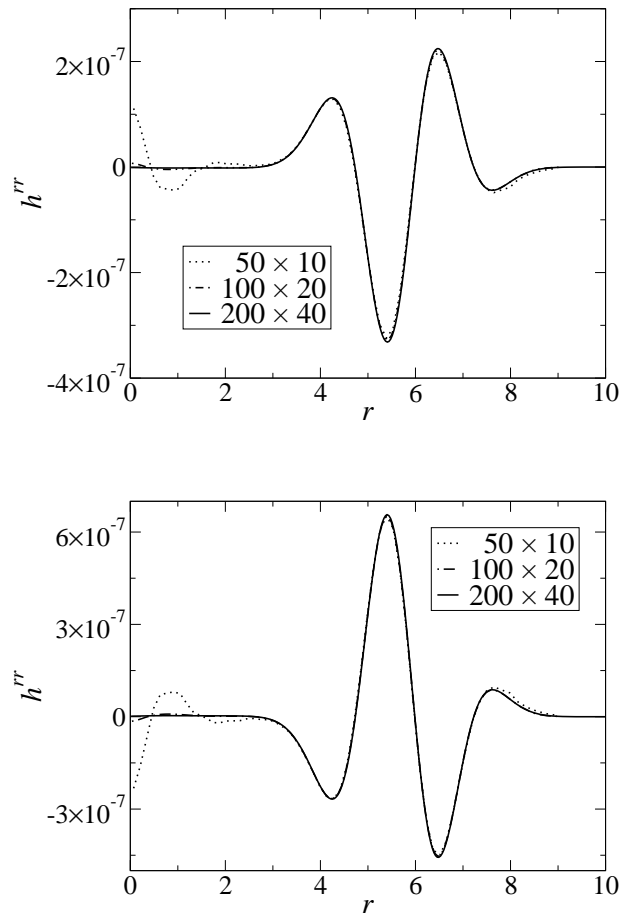

  \begin{center}
    \vspace*{0.4cm}
    \includegraphics[width=0.45\textwidth]{1a}\\
    \vspace*{1cm}
    \includegraphics[width=0.45\textwidth]{1b}
    \caption{
      Radial profile of $h^{rr}$ at $t=6$ at the equator (upper panel)
      and at the pole (lower panel). Three different resolutions 
      $n_r \times n_{\theta}$ are shown: $50 \times 10 $ (dotted lines), 
      $100 \times 20 $ (dot-dashed lines) and $200 \times 40 $ (solid lines).
    }
    \label{fig:Teu-wave}
  \end{center}
\end{figure}

\begin{figure}
\begin{center}
\includegraphics[width=0.45\textwidth]{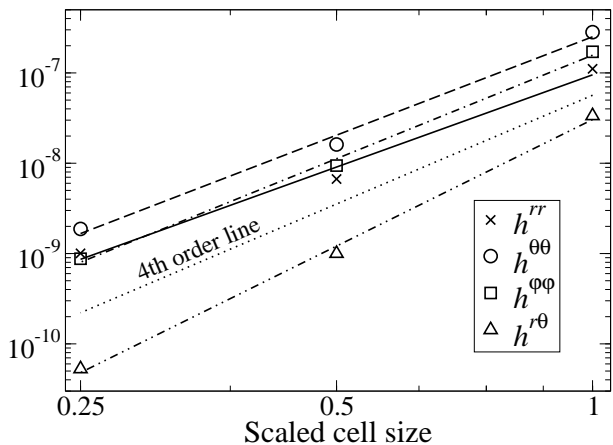}
\caption{Absolute errors of the numerical solutions with respect to the 
analytical values of all the non zero components of the tensor $h^{ij}$ at 
$(t,r,\theta)=(6,0,\pi/2)$ in terms of the (scaled) cell size. Solid, dashed, 
dot-dashed and dot-dot-dashed lines fit the errors for the component $h^{rr}$, 
$h^{\theta\theta}$, $h^{\varphi\varphi}$ and $h^{r\theta}$, respectively. 
Dotted line is the reference of fourth-order of convergence.}
\label{fig:Teu-conv}
\end{center}
\end{figure}

We display in Fig.~\ref{fig:Teu-wave} the radial profile of the component 
$h^{rr}$ at the end of the simulation, $t=6$, for three different numerical 
resolutions $n_r \times n_{\theta}$. Since the amplitude of the wave is 
sufficiently small to be considered a linear perturbation, we can compare the 
numerical solution with the analytical expression for the Teukolsky wave at 
each time. The solution agrees in the propagation speed of the wave, its 
amplitude and the asymptotic decay with increasing radius with the analytical 
solution. The Sommerfeld condition at the outer boundary produces ingoing
reflections with at most an amplitude square of the outgoing wave, as it is 
prescribed by the imposed condition. In Fig.~\ref{fig:Teu-conv} we plot the 
absolute errors of the numerical solution with respect to the analytical values
of all the non zero components of the tensor $h^{ij}$ at 
$(t,r,\theta)=(6,0,\pi/2)$ (Fig.~\ref{fig:Teu-wave} shows that the maximum of 
the absolute error appears at $r=0$), for different resolutions. We obtain an 
order of convergence of $3.4$, $3.6$, $3.8$, and $4.6$ for the components 
$h^{rr}$, $h^{\theta\theta}$, $h^{\varphi\varphi}$ and $h^{r\theta}$, 
respectively, which is close to the fourth-order of the corresponding 
Runge-Kutta method. Note that, since the background has $\beta^i=0$, the source
term $S_{w2}=0$ in Eq.~(\ref{e:evol_w_sources}); in this case we observe no 
significant reduction of the convergence order.

%%%%%%%%%%%%%%%%%%

\section{Evolution of equilibrium rotating neutron stars}

To test the performance of the passive FCF formulation in an astrophysical 
scenario we perform simulations of the evolution of isolated neutron stars. In 
this case the gravitational field is sufficiently strong to need to go beyond 
the Newtonian limit and at the same time the GW back-reaction is sufficiently 
small for the passive FCF approximation to be valid. Table~\ref{tab:approx} 
summarizes the approximations made in the simulations of this section.

\subsection{Initial model and grid}

We construct the initial data using the numerical code {\it rotstar\_dirac} of 
the \textsc{lorene} library~\cite{Lorene}, which computes axisymmetric and 
uniformly rotating neutron stars in equilibrium in the FCF formalism 
\cite{lapming06}. We use a polytropic equation of state, $P=K \rho^{\Gamma}$, 
with $\Gamma=2$ and $K=100$ (in $c=G=M_{\odot}=1$ units), to construct a 
neutron star with a 550 Hz rotation frequency, an ADM mass 
$M_{\rm ADM} = 1.4874\,M_\odot$  and a radius $R = 15.18$~km. The surface is 
located at a coordinate radius $r_* = 12.86$~km at the equator. The wavelegth 
of GWs at the frequency of the fundamental f-mode ($1.65$~kHz as measured from 
our numerical simulations of Sec.~\ref{sec:pertNS}) is 
$\lambda_{\rm F} = 181$~km. Following \cite{Thorne80} the (local) wave zone for
our neutron star model is located at 
$r >> \lambdabar \equiv \lambda / (2 \pi) = 28.8$~km. 

\begin{figure}
\begin{center}
\includegraphics[width=0.5\textwidth]{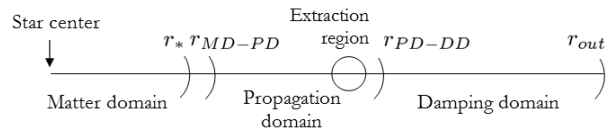}
\caption{Scheme of the radial grid used in the code for rotating neutron star 
simulations.}
\label{fig:grid}
\end{center}
\end{figure}

Thanks to the use of spherical coordinates, we can adapt the radial grid to
cover different domains of the space with the resolution needed on each domain.
In the case of a neutron star we have different resolution requirements inside 
the neutron star, where hydrodynamic variables have to be properly resolved, 
and outside, where it is sufficient to resolve the wavelength of the outgoing 
GWs.

\begin{table}
\begin{tabular}{ccc}
	\hline & regular resolution & high resolution \\ \hline
        $r_{\rm MD-PD} / r_*$                     &   1.19 & 1.19 \\
        $r_{\rm PD-DD} / \lambdabar_{\rm F}$       &   10.5 & 10.5 \\
        $r_{\rm out} / \lambdabar_{\rm F}$         &   104.2 & 104.2 \\
        $\Delta r_{\rm PD-DD} / \lambda_{\rm F}$    &   5 & 10 \\
        $n_{r, {\rm MD}}$                          &   80 & 160 \\
        $n_{r, {\rm PD}}$                          &   346 & 690 \\
        $n_{r, {\rm DD}}$                          &   43 & 88 \\  
        $n_{r}$                                   &   469 &  938 \\ 
        $n_{\theta}$                               &   16 & 32 \\ \hline
\end{tabular}
\caption{Parameters used in the finite differences grid in simulations of 
neutron star evolution.}
\label{tab:resolution}
\end{table}

For the finite difference grid, we consider three radial domains covering the 
computational domain (see Fig.~\ref{fig:grid}): the {\it matter domain} (MD) 
contains the neutron star, extends from the center to a radius $r_{\rm MD-PD}$ 
slightly larger than the stellar radius $r_*$. This domain is covered by an 
equidistant radial grid. The {\it propagation domain} (PD), extends from 
$r_{\rm MD-PD}$ to a radius $r_{\rm PD-DD} >> \lambdabar$. In this region the 
radial grid spacing increases geometrically outwards, such that the GWs are 
well resolved. Near the outer edge of the domain, the GWs reach the wave zone, 
and hence it is an appropriate radius to perform the GW extraction. The 
{\it damping domain} (DD) extends from $r_{\rm PD-DD}$ to the outer boundary of
the numerical grid $r_{\rm out}$. We locate the outer boundary such that an 
outgoing wave generated at the center at $t=0$ and traveling at the speed of 
light reaches the outer boundary at the end of the simulation. This 
configuration minimizes the effect of spurious numerical artifacts at the outer
boundary which where found in our preliminary work \cite{CC10}. In the damping 
domain the radial grid spacing increases geometrically outwards and the 
wavelength of the GWs is not well resolved. This produces an effective damping 
of the outgoing GWs which helps to reduce the effect of the outer boundary 
conditions. To construct the finite difference grid we need to provide the 
values of $r_{\rm MD-PD}$, $r_{\rm PD-DD}$, $r_{\rm out}$, the number of grid 
points inside of the MD, the grid spacing $\Delta r_{\rm PD-DD}$ at 
$r_{\rm PD-DD}$ (which automatically fixes the number of points inside PD), and
the cell radial spacing ratio between consecutive ones at DD (which fixes the 
number of points in this domain). We perform simulations with two resolutions, 
labeled regular and high, whose grid parameters are given in 
table~\ref{tab:resolution}. 

For the spectral grid we use two resolutions labeled regular and high. The 
regular resolution grid consists of $5$ radial spectral domains covering the 
finite difference grid, $4$ domains with $33$ and $1$ with $17$ radial 
collocation points, and a compactified spectral domain from $r_{\rm out}$ to 
infinity with $17$ collocation points. The $\theta$ direction is covered by $5$
collocation points. The high resolution grid consist of $13$ spectral domains, 
$6$ domains with $33$ and $7$ with $17$ collocation points, and a compactified 
domain with $17$ collocation points. The $\theta$ direction is covered by $17$ 
collocation points.

\subsection{Equilibrium neutron stars}
\label{sec:eq_neu_stars}

\begin{figure}
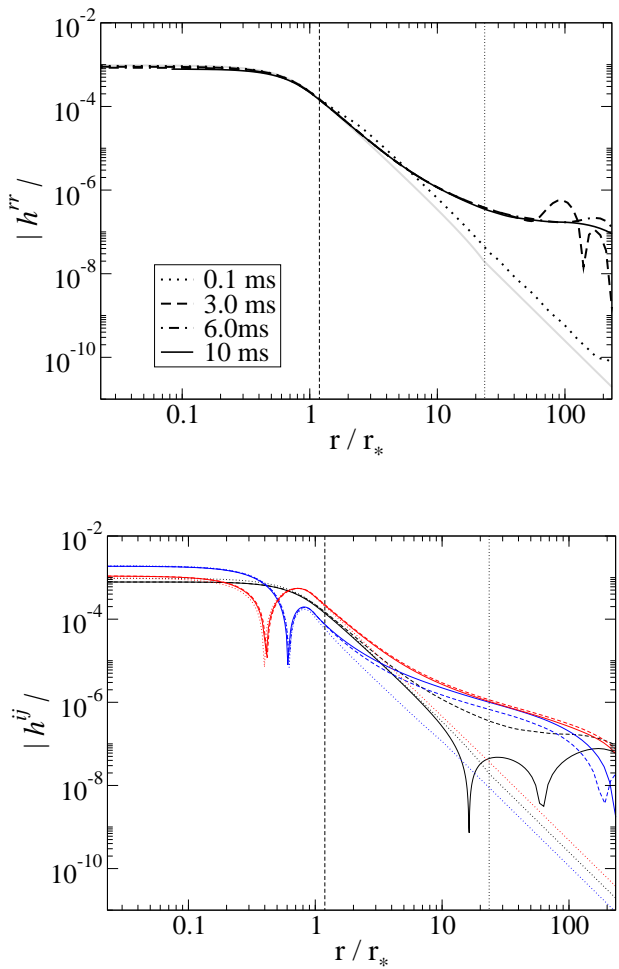

\begin{center}
\includegraphics[width=0.45\textwidth]{4a}
\vspace*{0.8cm} \\
\includegraphics[width=0.45\textwidth]{4b}
\caption{Radial profiles of $h^{ij}$ for numerical simulations of neutron
stars. The upper panel shows four snapshots of the evolution of $h^{rr}$
and the initial stationary solution (grey line). The lower panel shows the 
components $h^{rr}$ (black lines), $h^{\theta\theta}$ (blue lines), and 
$h^{\varphi\varphi}$ (red lines) for the initial stationary configuration 
(dotted lines), and the final configuration at $10$~ms (dashed lines for the 
regular and solid lines for the high resolution spectral grid).}
\label{fig:sta-cfc}
\end{center}
\end{figure}

Before attempting to solve the full coupled evolution of spacetime and 
hydrodynamics in neutron stars, we perform simulations in a more simplified 
setting. We evolve the hyperbolic sector of the passive FCF formalism in a 
fixed non-trivial background (non-vanishing $N$, $\beta^i$, $\psi$ and 
hydrodynamic variables) corresponding to the equilibrium configuration of a 
rotating neutron star. Therefore, we evolve stationary initial data for the 
variable vector ${\bf W}$ and keep the variable vector ${\bf U}$ and ${\bf V}$ 
fixed during the evolution of ${\bf W}$. 

Our fiducial model has regular resolution concerning both the finite difference
grid and the spectral one. The background model is computed with FCF gravity, 
but we recompute the elliptic part, ${\bf V}$, at the beginning with the CFC 
approximation, i.e., the background is evolved in the same way as it is in the  
simulations of the next section, where the vector ${\bf V}$ is also evolved in 
time. The consequences of this modification on the background metric are 
evaluated below. We evolve the system of equations for the vector ${\bf W}$ for
$10$~ms. Due to small numerical discrepancies between the initial data for 
${\bf W}$ and the numerical stationary solution of the equations a perturbation
in the vector ${\bf W}$ appears and propagates outwards. The upper panel of 
Fig.~\ref{fig:sta-cfc} shows four snapshots of the evolution of the $h^{rr}$ 
component compared to the stationary solution (grey line). Note that the 
perturbation reaches the outer boundary at about $3$~ms, well before the end of
the simulation. This is due to the unphysical superluminical propagation of the
wave in the damping domain, where its wavelength is unresolved. However, due to
the smallness of $h^{ij}$ at the outer boundary, about $3$ orders of magnitude 
smaller than at the center, spurious reflections are not noticeable in the 
simulations. At the end of the simulation the outgoing wave leaves the 
numerical domain and an equilibrium configuration remains. We can compare this 
solution with the initial stationary data to look for numerical discrepancies. 
At the center the initial configuration is recovered within $\sim 10\%$ 
accuracy. For distances $r/r_* \geq 10$ (see upper panel of 
Fig.~\ref{fig:sta-cfc}) there are larger deviations, and the components of 
$h^{ij}$ decay approximately as $r^{-1}$ instead of $r^{-3}$ as in the initial 
stationary model. Since GWs are contained in the part of $h^{ij}$ decaying as 
$r^{-1}$, the erroneous decay observed in the simulations at large distances 
will lead to a constant offset in the computed GW amplitude. This offset can be
comparable to the amplitude of GWs produced by small perturbations 
(see~\cite{CC10} for more details). We discuss the possible causes for this 
behavior below.

\begin{figure*}
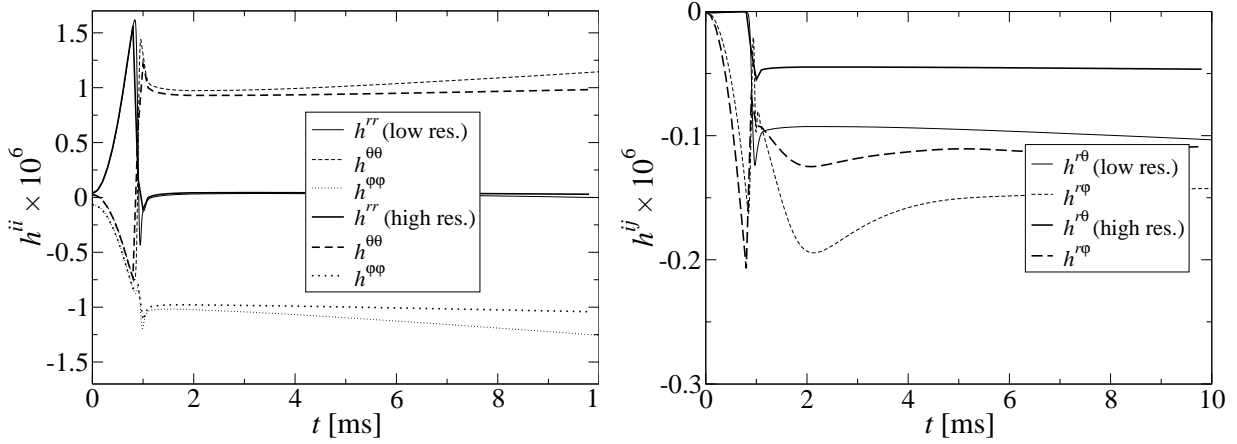

\begin{center}
\includegraphics[width=0.45\textwidth]{5a} 
\includegraphics[width=0.45\textwidth]{5b}
\caption{Time evolution of $h^{ij}$ 
at $r/r_{*}=19.44$ and $\theta=\pi/2$. Left (right) panel shows the diagonal 
(non-diagonal) components. We plot two different finite differences 
resolutions: regular (thin lines) and high (thick lines).}
\label{fig:sta-conv}
\vspace*{0.1cm}
\end{center}
\end{figure*}

\begin{figure*}
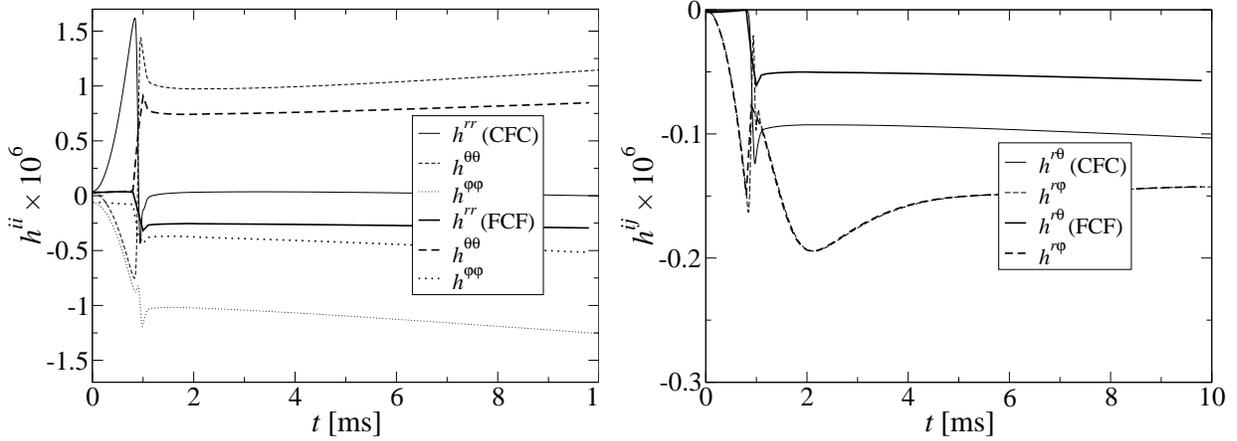

\begin{center}
\includegraphics[width=0.45\textwidth]{6a} 
\includegraphics[width=0.45\textwidth]{6b}
\caption{Same as Fig.~\ref{fig:sta-conv}, but we compare simulations using the 
CFC approximation (thin lines) and the full FCF metric (thick line) for the 
background metric {\bf V}.}
\label{fig:sta-conv-fcf}
\end{center}
\end{figure*}

One important reason for the appearance of offsets is the accuracy of the 
numerical solution of the elliptic equations. We have performed simulations 
increasing the spectral grid resolution, but keeping the same regular 
resolution for the finite difference grid. The spectral grid resolution affects
the accuracy of the computation of the background ${\bf V}$ computed at the 
beginning of the simulation. The lower panel of Fig.~\ref{fig:sta-cfc} shows 
the final configuration of the diagonal components of $h^{ij}$ for the regular 
and high resolution spectral grid, compared to the initial equilibrium 
configuration. The error at the center of the equilibrium configuration at the 
end of the simulation does not improve with respect to the regular resolution 
spectral grid. The erroneous decay of $h^{ij}$ improves significantly for the 
$h^{rr}$ component, and we recover the correct $r^{-3}$ decay in the whole 
propagation domain. However, the $h^{\theta\theta}$ and $h^{\varphi\varphi}$ 
components do not improve significantly. Therefore, there is a strong 
sensitivity of the $h^{rr}$ component on the spectral metric resolution, and 
hence on the accuracy of the computation of ${\bf V}$, because the variables 
in ${\bf V}$ appear in the leading terms of the equations for ${\bf W}$, 
Eqs.~(\ref{e:evol_h})--(\ref{e:evol_w}). However, spectral resolution does not 
seem to cure the problems in $h^{\theta\theta}$ and $h^{\varphi\varphi}$ which,
as we show below, are related to other sources of inaccuracy in the solution of
${\bf V}$. Because of the better performance of the high resolution spectral 
metric we use this resolution for all simulations hereafter.

To check the effect of the finite difference grid resolution we have performed
simulations with the high resolution grid, keeping the high resolution spectral
grid fixed. The finite difference grid resolution affects the accuracy of the 
solution of the vector ${\bf W}$, i.e. the evolution of $h^{ij}$. In 
Fig.~\ref{fig:sta-conv} we plot the time evolution of the components of the 
tensor $h^{ij}$ (left panel for the diagonal components and right panel for 
non-diagonal components) at $r/r_{*}=19.44$, for the two finite difference grid
resolutions. This coordinate radius is close to the outer boundary of the 
propagation domain and the inner edge of the extraction region for GWs (see 
next section). About $1$~ms after the beginning of the simulation, the outgoing 
wave reaches this radius visible in Fig.~\ref{fig:sta-conv} as a 
sudden rise of all components of $h^{ij}$. After the outgoing wave leaves the 
numerical domain the value of $h^{ij}$ does not settle down to the initial 
equilibrium value, but to an offset value, decaying as $r^{-1}$. All components 
converge with finite difference grid resolution to an offset value.
The offset of the component $h^{rr}$ cannot be appreciated
in Fig.~\ref{fig:sta-conv} since it is much smaller than in the other components.

Apart from the numerical error in the computation of ${\bf V}$ due to a finite 
spectral grid resolution, the approximating of the vector ${\bf V}$ by the 
solution of the CFC equations instead of the full FCF elliptic equations might 
also introduce small errors in the vector ${\bf V}$, which are sufficiently 
large to explain the observed offsets. Although we still cannot solve the FCF 
elliptic equations with CoCoNuT in a simulation with spacetime evolution, for 
the case of fixed vectors $\bf U$ and $\bf V$ considered in this section, we 
can take the full FCF solution for $\bf V$ computed by the initial data solver 
{\it rotstar\_dirac}. We have performed simulations with the regular and high 
resolution finite difference grids and the FCF background metric vector 
$\bf V$. The resolution used for the spectral solver in the initial data 
generator, {\it rotstar\_dirac}, fixes the numerical accuracy of the vector 
${\bf U}$. Unfortunately, due to internal code limits of {\it rotstar\_dirac}, 
the maximum resolution that we could achieve was $8$ radial domains, $5$ with 
$33$ and  $3$ with $17$ collocation points, and $17$ collocation points in the 
$\theta$ direction. This resolution lays in between the regular and the high 
resolution spectral grids used for CFC metric computations. 
Fig.~\ref{fig:sta-conv-fcf} compares the evolution of $h^{ij}$ close to the GW 
extraction radius, $r/r_{*}=19.44$, with the CFC and the FCF background. The 
offset in $h^{\theta\theta}$, $h^{\varphi\varphi}$ and $h^{r\theta}$ is reduced
when the CFC approximation is removed and the background is computed with the 
full general relativistic FCF formulation. The offset in $h^{rr}$ increases, 
however, although this is expected since the resolution of the spectral grid is
lower in the FCF case than in the CFC case. We observe no change in the offset 
of the $h^{r\varphi}$ component.

We conclude that the main reason for the offset in $h^{ij}$ at large distance 
from the source is the accuracy of the computation of the vector ${\bf V}$. 
Both the resolution of the spectral grid and the neglected terms in the 
elliptic part due to the passive FCF approximation are responsible for this 
loss of accuracy, affecting in each case different components of $h^{ij}$. This
defines the spectral grid resolution necessary for the simulations in the next 
section, but the use of the passive FCF approximation still introduces an 
offset which cannot be removed. The resolution tests show that the finite 
difference grid is adequate to evolve ${\bf W}$, and is not responsible for the
offset.

The order of convergence of the code is difficult to evaluate due to the fact 
that if we increase the finite difference grid resolution the solution 
converges towards an offset solution and not to the equilibrium one, and at the
same time this offset converges with the spectral grid resolution and is 
affected by the passive FCF approximation. However, we can check the time 
behavior of the stationary solutions after the outgoing wave is gone. In
Fig.~\ref{fig:sta-conv} the time evolution of all $h^{ij}$-components suffers a
time-drift which is due to numerical diffusion.
We fit the evolution of the $h^{\theta\theta}$-component between
$t_1 = 4$~ms and $t_2=10$~ms to $h^{\theta\theta}(t_1) + C \cdot (t_2-t_1)^p$. 
The fitted value for the power $p$ is $1.7$ and $1.8$ for the regular and high  
resolution simulations, respectively. If we consider that $C\sim {\Delta r}^p$,
we can also compute the power as 
$p = log(C_{\rm regular}/C_{\rm high}) / log {\,2}$, which is $1.96$. 
Therefore, the order of convergence is close to second-order, due to the 
mixture of implicit and explicit terms in the fourth-order Runge-Kutta scheme. 
If we apply this analysis to other components of $h^{ij}$, we find a similar 
order of convergence.

%%%%%%%%%%%%%%%%%%

\subsection{Perturbed equilibrium configuration of rotating neutron star}
\label{sec:pertNS}

\begin{figure*}
\begin{center}
\includegraphics[width=0.95\textwidth]{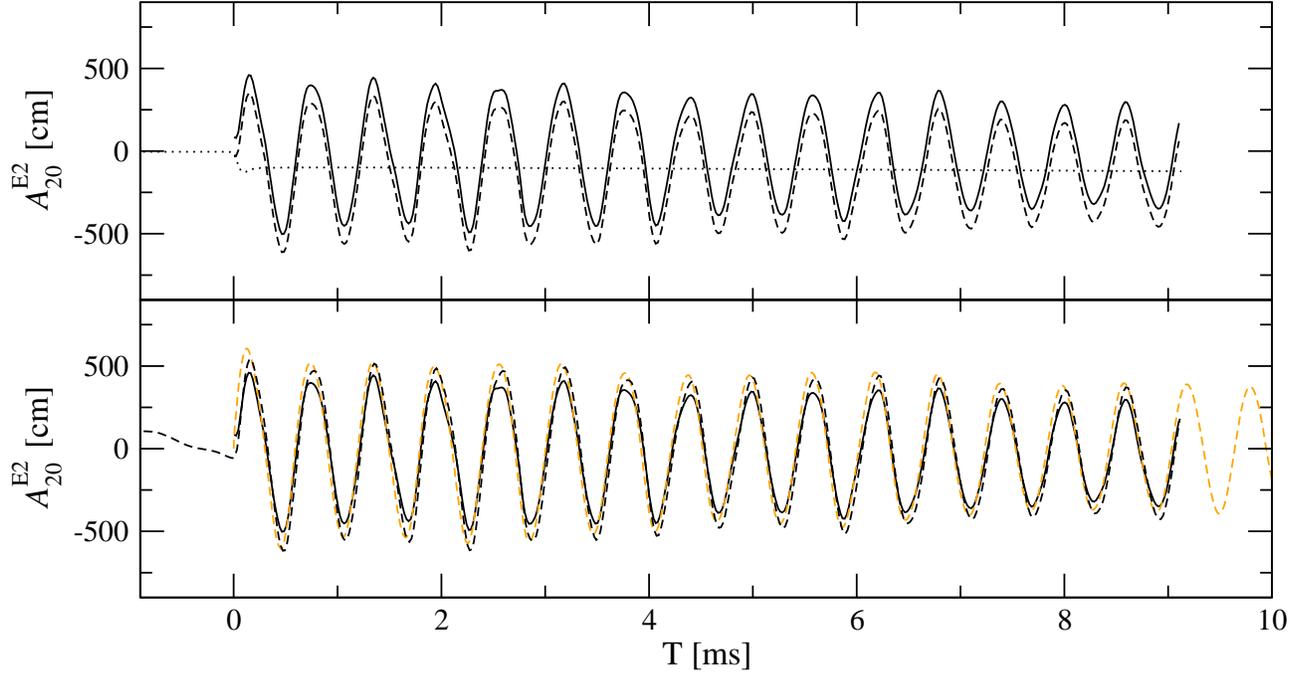}
\caption{Quadrupolar component of the GW extracted from simulations of an 
oscillating neutron star. The upper panel shows the extrapolated waveform at 
infinity (black dashed line) and the offset-corrected waveform extrapolated at 
infinity (black solid line). The offset is corrected using the average value of
the waveform of a non-perturbed neutron star simulation (black dotted line). 
The lower panel shows the offset-corrected waveform at an extraction radius of 
$r/r_{*}=19.44$ (black dashed line) and extrapolated at infinity (black solid 
line), compared to the quadrupole formula (orange dashed line).}
\label{fig:perturbed-1}
\end{center}
\end{figure*}

The last test consists in the evolution of an oscillating neutron star with 
coupled hydrodynamics and spacetime evolution, i.e. we evolve the coupled 
system for ${\bf U}$, ${\bf V}$ and ${\bf W}$ in the passive FCF approximation.
We use the regular resolution finite difference grid and the high resolution 
spectral grid. We initiate the oscillations by adding a small $l=2$ velocity 
perturbation (about $1\%$ of the speed of light) to the stationary initial data
used in the previous subsection.  Previous preliminary studies~\cite{CC10} show
that the gravitational radiation has to be extracted close to $r_{\rm PD-DD}$ 
but still inside the propagation domain, where the wavelength of the 
fundamental mode is resolved by about 5 grid points in the regular resolution 
finite difference grid. 

For asymptotically flat spacetimes, as the one used in our simulations, and 
since the Dirac gauge tends to the TT gauge far from the strong field region 
\cite{Smarr78}, it is possible to compute the amplitude of the plus 
polarization of the GW, i.e. the real part of the $\Psi_4$ Weyl scalar, 
detected by an observer at a distance $R$ with an observation angle $\Theta$ 
with respect to the rotation axis as
\be
	h_{+} (R, \Theta, T) = \lim_{R\to\infty} \frac{h^{\varphi\varphi} (R, \Theta, T) 
- h^{\theta\theta} (R, \Theta, T)}{2}.
\ee
In axisymmetry, the cross polarization, $h_{\times}$, vanishes. In general, one
should compute outgoing null geodesics for each angle $\theta$ and determine 
the observation angle $\Theta$ and the distance $R$, and then use the numerical 
value of $h^{ij}$ at the extraction radius to determine $h_{+}(R, \Theta, T)$. 
Thanks to the equatorial symmetry, the curves $\theta=\pi/2$ on $\Sigma_t$ are 
null geodesics, and will be observed at a distance $R$ with inclination angle 
$\Theta=\pi/2$. The distance to an observer located at a coordinate radius $r$ 
can be easily computed as
\be
	R (r)= \int_0^{r} \sqrt{\gamma_{rr}(r,\theta=\pi/2)}\, dr 
\approx \int_0^r \psi(r,\theta=\pi/2)^2\,dr.
\label{e:R}
\ee

For our neutron star model, the spacetime surrounding it is not extremely 
curved, and radial null geodesics at other angles are approximately curves with
constant $\theta$, i.e. $\Theta \approx \theta$. If we integrate the distance 
$R$ as in Eq.~(\ref{e:R}) at different angles $\theta$, the difference between 
equator and polar axis is about $0.04\%$. Hence, for neutron stars we can 
safely compute $h_{+}$ at any angle using the numerical value of $h^{ij}$ at 
the extraction radius $r_{\rm ext}$ as
\be
	h_{+} (R, \theta, T)\approx \frac{h^{\varphi\varphi}(r_{\rm ext}, \theta,
t)-h^{\theta\theta}(r_{\rm ext}, \theta, t)}{2} \frac{R(r_{\rm ext})}{R},
\label{eq:gw_eq}
\ee
where $T \equiv t - R / c$ is the retarded time. Note that Eq.~(\ref{eq:gw_eq})
is approximate and valid only in the wave zone, i.e. for 
$r_{\rm ext} >> \lambdabar_{\rm F}$.

In axisymmetry and with equatorial symmetry, the multipolar decomposition of 
the radiation field is \cite{Thorne80}:
\begin{equation}
h_{+} (R, \theta, T) = \frac{1}{R} \left ( 
A^{E2}_{20}(T)  T^{E2, 20} (\theta) + A^{E2}_{40}  T^{E2, 40} (\theta) +
... \right )
\end{equation}

We have extracted the amplitude of the $l=2$ and $4$ multipoles, $A^{E2}_{20}$ 
and $A^{E2}_{40}$, respectively, from the numerically computed $h_{+}$, 
at all radial points between $15 r_*$ and $19.44 r_*$. For 
post-Newtonian sources, as in our case, we expect the amplitude of the 
multipoles to decrease with $l$ \cite{Thorne80}. We were unable to extract 
higher-order multipoles because of the smallness of the signal, which made the 
numerical extraction too noisy. Since the extraction radii are relatively close 
to the source, in the interval $6.7 \lambdabar$-$8.7 \lambdabar$, there is a 
small radial dependence of the GW amplitude and phase on the extraction radius. 
This dependence corresponds to $r^{-p}$, $p\geq2$, components in $h^{ij}$, 
where the leading term is $r^{-2}$ and can be fitted to extract the waveform at
infinity using a procedure similar to \cite{Baiotti09}. To extrapolate both 
phase and amplitude, we proceed in two steps:\\
i) First, we perform a least squares 
fit of the retarded time of each maxima in the waveform as a function of the 
extraction radius $R$ to 
\begin{equation}
T_{\rm max} (R) = T_{\rm max} (\infty) + \frac{C}{R}.
\end{equation}
We use the average value of $C$ for all maxima, $<C>/r_*=0.2245$~ms, to correct
the phase of the waveform as
\begin{equation}
T (\infty) = T (R) - \frac{<C>}{R}.
\end{equation}
ii) Once the phase is corrected, we fit the amplitude of the waveform at 
constant $T(\infty)$ as a function of $R$ to 
\begin{equation}
A^{E2}_{20} (R, T(\infty)) = A^{E2}_{20} (\infty,T(\infty)) + \frac{C'}{R}.
\end{equation}
The resulting fitted value of $A^{E2}_{20} (\infty,T(\infty))$ is the waveform 
extrapolated at infinity. To estimate the finite distance effects we present 
results of the waveform extracted at a finite distance 
$r_{\rm ext} = 19.44 r_*$ and extrapolated at infinity. 

An alternative approach to compute the GW amplitude is to use the 
post-Newtonian wave-generation formalism. This is possible if the sources allow
for a post-Newtonian expansion, i.e. $(v/c)^2\sim M/r_* < 1$. For slow-motion 
sources \cite{Thorne80}, for which $r_*<<\lambdabar$, it is possible to write 
the amplitudes of the different multipoles as volume integrals over the matter 
sources. Truncating the integrals at the lower post-Newtonian level, i.e. with 
Newtonian sources, the quadrupolar component results in the well known 
quadrupole formula \cite{Einstein18}. In axisymmetry the quadrupole formula 
reads
\be
A^{E2}_{20}= 8 \sqrt{\frac{\pi}{15}} \frac{d^2}{dt^2} \left \{
\int D \,(3 z^2 - 1) \,r^2 \sqrt{\gamma} \, dr
d\theta d\varphi)
\right\},
\ee
with $z\equiv \cos {\theta}$. Using the continuity equation, $J^\mu_{;\mu}=0$, 
one of the time derivatives can be removed analytically 
(cf. \cite{Blanchet90}),
\bea
A^{E2}_{20}= 8 \sqrt{\frac{\pi}{15}} \frac{d}{dt} \Bigg \{&&
\int D \,\Big( v^{* r} (3 z^2 - 1) \nonumber \\
&&- 3 v^{* \theta} z \sqrt {1-z^2}\,\, \Big )\,r \sqrt{\gamma} \, 
dr d\theta d\varphi) \Bigg \},
\eea
where $v^{* i}\equiv \alpha v^i - \beta^i$.
The latter formula is more convenient from the numerical point of view, since 
only one time derivative has to be evaluated numerically. We use it in this 
work.

\begin{figure*}
\begin{center}
\vspace*{0.1cm}
\includegraphics[width=0.95\textwidth]{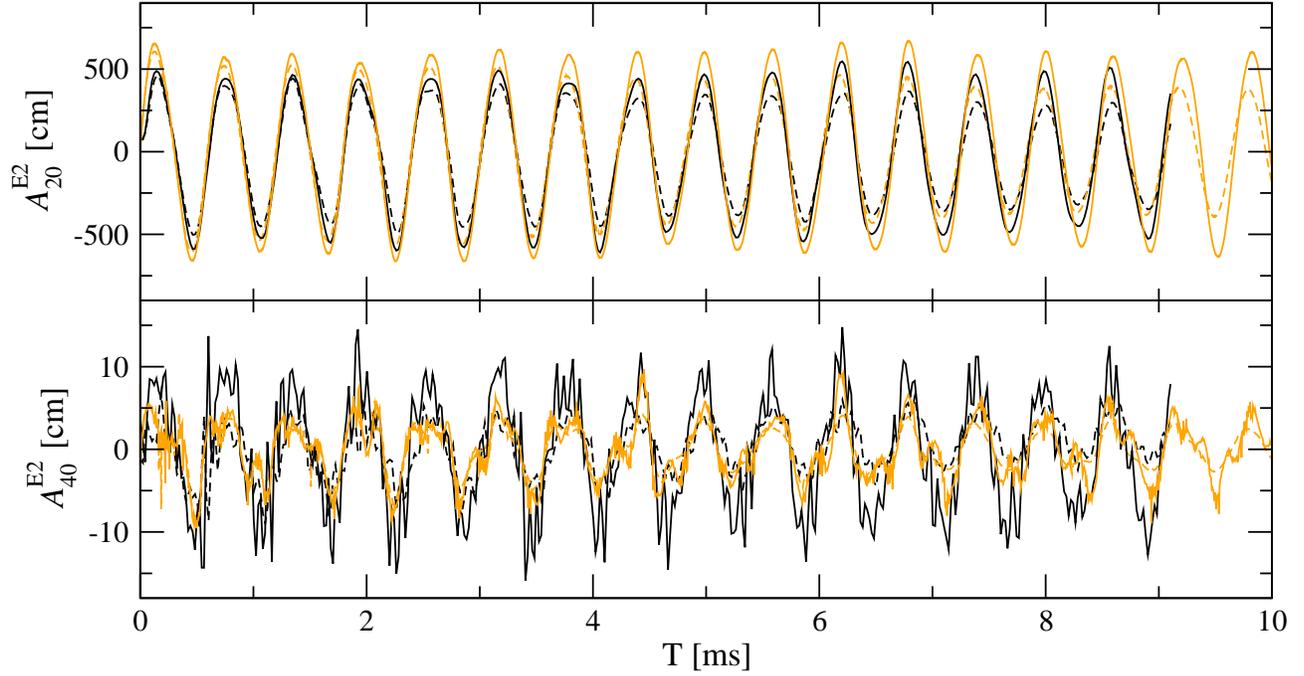}
\caption{GW extracted from simulations of an oscillating neutron star. Upper 
and lower panels show the quadrupolar and hexadecapolar component respectively.
On each panel we compare simulations for regular (dashed lines) and high (solid
lines) resolution finite differences grid. The offset-corrected and waveform 
extrapolated at infinity computed with the direct extraction method (black 
lines) is compared to the PN method (quadrupole and hexadecapole formulae, 
orange lines).}
\label{fig:perturbed-2}
\end{center}
\end{figure*}

In equatorial symmetry, the hexadecapolar component is \cite{Moenchmeyer91, 
Faye03}:
\be
A^{E2}_{40}= \frac{\sqrt{5 \pi}}{126} \frac{d^4}{dt^4} \left \{
\int D \,\left (7 z^4 - 6 z^2 + \frac{3}{5} \right) \,r^4 \sqrt{\gamma} \, dr
d\theta d\varphi) \right\}.
\ee
In a similar way as for the quadrupole formula, it is possible to remove one 
time derivative using the continuity equation (cf. \cite{Moenchmeyer91, 
Faye03}):
\bea
A^{E2}_{40}= &&\frac{2\sqrt{5 \pi}}{63} \frac{d^3}{dt^3} \Bigg \{ 
\int D \,\Bigg[
v^{*r} \left (7 z^4 - 6 z^2 + \frac{3}{5} \right) \nonumber \\
&& +v^{*\theta} \left( (3-7 z^2) z \sqrt{1-z^2}\right)
\Bigg ]\,r^3 \sqrt{\gamma} \, dr d\theta d\varphi) \Bigg\}.
\eea

\begin{figure}
\begin{center}
\vspace*{0.1cm}
\includegraphics[width=0.45\textwidth]{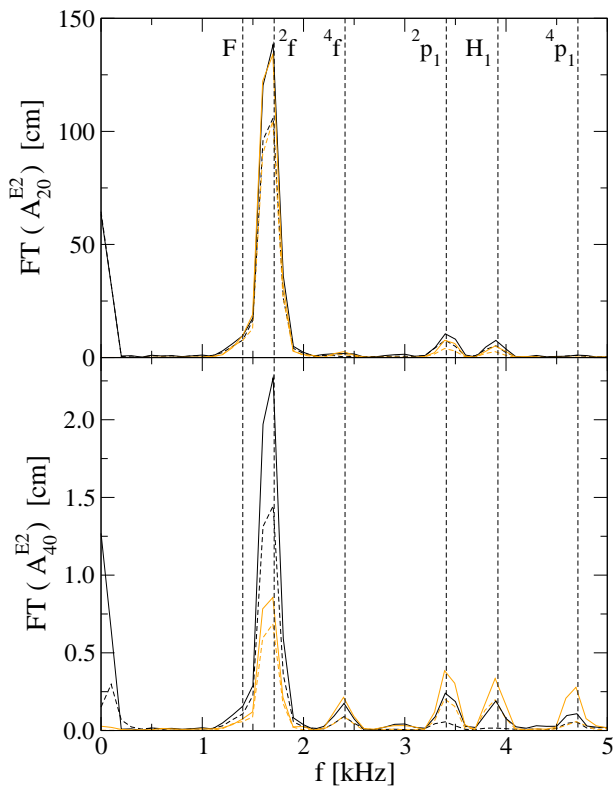}
\caption{Fourier transform of the GW amplitude of an oscillating neutron star. 
Upper and lower panels show the quadrupolar and hexadecapolar component 
respectively. On each panel we compare simulations for regular (dashed lines) 
and high (solid lines) resolution finite differences grid. The Fourier 
transform of the offset-corrected waveform computed with the direct extraction 
method (black lines) is compared to the quadrupole formula (orange lines).}
\label{fig:fftperturbed-2}
\end{center}
\end{figure}

In the upper panel in Fig.~\ref{fig:perturbed-1}, we plot the time evolution of
the quadrupolar component $A^{E2}_{20}$ computed with Eq.~(\ref{eq:gw_eq}) 
({\it direct extraction} hereafter) extrapolated at infinity. The waveform 
clearly shows a constant offset which is unphysical. The dotted line shows the 
GW corresponding to the model in the previous section with the same resolution,
but fixed ${\bf U}$ and ${\bf V}$, and no initial perturbation. In this case we
do not observe any oscillation, since the star itself is not oscillating, but 
an offset appears of similar magnitude as in the oscillating case. We can use 
the value of the offset at each time from the non-perturbed simulation to 
remove the offset in the oscillating simulation by simple subtraction of both 
GW signals.

In the lower panel of Fig.~\ref{fig:perturbed-1} we compare the offset 
corrected waveform to the result of the quadrupole formula. There is a 
remarkable good agreement in the frequency and phase. The Fourier transform of 
the waveform is shown in the upper panel of Fig.~\ref{fig:fftperturbed-2}. 
Within the numerical frequency resolution of the Fourier transform of the 
waveform, about $0.02$~kHz, we do not observe differences in the frequency 
between the direct extraction method and the quadrupole formula. That sets an 
upper limit of $1\%$ for the frequency difference in the fundamental 
oscillation mode, $f_{\rm F}=1.66$~kHz. The phase difference between both GW 
extraction methods, estimated as the relative difference in the retarded time 
at the maxima, is smaller than $1\%$. The corrected signal (solid line in 
Fig.~\ref{fig:perturbed-1}) agrees with the quadrupole formula within $30\%$. 
Therefore, the only big discrepancy with the quadrupole formula is due to the 
error committed in the computation of the vector ${\bf V}$, using the 
passive FCF approximation and the spectral grid resolution. The quadrupole 
formula is an approximate formula, which is valid in the slow-motion 
post-Newtonian limit. The error in the formula should be of the order 
$(v/c)^2\sim M/r_*$, which for our system is $\sim 17\%$. Therefore, the $30\%$
discrepancy in the amplitude is compatible with the approximation error of the 
quadrupole formula. Note that the waveform extracted at finite distance suffers
from additional errors. For example, at $19.44 r_*$ (black dashed line in 
Fig.~\ref{fig:perturbed-1}) the amplitude of the waveform differs about $30\%$ 
from the extrapolated waveform at infinity and its phase about $5\%$.

In the case of the hexadecapolar component $A^{E2}_{40}$ (lower panel of
Figs.~\ref{fig:perturbed-2} and \ref{fig:fftperturbed-2}), the direct 
extraction method shows an important contribution due to numerical noise. Note 
that the hexadecapolar component is about a $2\%$ contribution to the total 
waveform $h_+$. Therefore, this numerical noise appears because of errors in 
the evolution of $h^{ij}$ below $1\%$, which are expected in our simulations. 
In this case the hexadecapole formula provides a good estimate of the phase 
and the frequency ($\sim 5\%$), however, the error in the amplitude is about 
$50\%$. Note that for the hexadecapolar component there are possible sources of
error in both the direct extraction method, due to the smallness of the 
amplitude, and in the hexadecapole formula, due to the three numerical time 
derivatives that we have to perform. Therefore, it is difficult to disentangle 
which one is a better approximation to the waveform. Nevertheless, both methods
provide a reasonable agreement, and we are confident that the amplitude 
computed with any of the two methods is a rather good order-of-magnitude 
estimate of the hexadecapolar component.

To test the effect of the finite difference grid resolution, we compare 
simulations with regular and high resolution, and the same high spectral metric
resolution. Fig.~\ref{fig:perturbed-2} shows the offset-corrected values of 
$A^{E2}_{20}$ and $A^{E2}_{40}$ for both resolutions. For the offset correction
we use the corresponding regular and high resolution simulations of the 
previous section. We compare with the post-Newtonian wave generation formalism 
(quadrupole and hexadecapole formulae, {\it PN method} for short) for both 
resolutions. In both methods, direct extraction and PN method, we observe a 
damping of the waveforms which is reduced when increasing the resolution. The 
PN method only uses information in the GW generation zone ($r<r_*$), contrary 
to the direct extraction method, where the wave is propagated from the 
generation zone to the wave zone. However, in both cases the damping observed 
in the waveforms is of similar magnitude. That means that the source of the 
damping must be caused by numerical inaccuracies in the region close to the 
star, but not in the propagation domain. In other words, we observe a numerical
damping of the oscillations of the star itself, but not of the waves during 
their propagation towards the GW extraction point.

%%%%%%%%%%%%%%%%%%%%%%%%%%%%%%%%%%%%%%

\section{Summary and conclusions}

We have reviewed the fully constrained formalism, which is a natural 
generalization of the CFC approximation of GR, and we have expressed the system
of FCF equations in a form suitable for numerical simulations. We have 
presented a numerical scheme to solve the FCF system using spectral methods for
the elliptic part and finite difference schemes for the hyperbolic part. In the
simulations presented here we have neglected the back-reaction of the GWs onto 
the dynamics, which we call passive FCF. This work focuses on the stability and
convergence of the hyperbolic part of the FCF equations, since the stability 
issues of the elliptic part were considered by \cite{CC09}. We have presented a
fourth-order finite difference scheme to solve the system of hyperbolic 
equations that makes use of implicit relations, to provide the necessary 
stability of the algorithm. We have solved the equations in spherical 
coordinates and axisymmetry.

We have performed convergence tests for the hyperbolic part, in which the 
gravitational radiation of the system is encoded, using a simple vacuum test 
with known analytical solution: the Teukolsky waves. We have shown the 
stability and convergence of the numerical evolution which is consistent with 
the fourth-order convergence of the numerical scheme. We have performed the 
evolution of equilibrium neutron stars and checked that the numerical code is 
able to maintain such configurations in equilibrium keeping the hyperbolic part
to an accuracy of second-order. We interpret the drop to second-order 
convergence in all our simulations with matter content as an inconsistency in 
our numerical scheme in this regime, due to the mixture of explicit and 
implicit terms in the Runge-Kutta scheme. In order to improve the order of 
convergence one should use IMEX methods to solve the system of equations 
\cite{AsRuuSpi, PaRu}. Although this approach is beyond the scope of this work,
it may be considered in the future.

Finally, we have performed simulations of the evolution of an oscillating 
equilibrium neutron star. We have extracted the GW signature from the metric 
components in the wave zone. We have compared the results from the direct 
extraction method with calculations using the post-Newtonian wave generation 
formalism, namely the Newtonian quadrupole and hexadecapole formulae. We found 
that the approximate quadrupole formula describes within $\sim 30\%$ error 
the quadrupolar component of the wave. This is consistent with its nominal 
post-Newtonian error ($\sim 20\%$). Similar good agreement of the quadrupole 
formula with BSSN simulations was found by \cite{Shibata03,Reisswig11}. We were
able to extract the hexadecapole component of the wave, although numerical 
noise is considerably larger than in the quadrupole component. The comparison 
with the Newtonian hexadecapole formula agrees in frequency, phase and order of
magnitude, however the comparison is limited by the numerical accuracy of both 
wave extraction methods. We found that the evolution of the hyperbolic part of 
the metric is very sensitive to inaccuracies in the elliptic sector, resulting 
in offsets in the GW signature. The main sources for inaccuracies are the 
number of collocation points in the spectral solver for the elliptic part and 
the absence of back-reaction terms. We were able to get convergent results 
increasing the spectral resolution, however, an offset still remained due to 
the passive FCF approximation. We conclude that GWs back-reaction should be 
included in the future, as well as an improvement in the accuracy of the 
numerical solution of the elliptic equations, in order to remove these 
offsets. 

We note that all simulations that we have performed in this work make use of 
spherical orthonormal coordinate grids, for the whole computation domain, in 
axisymmetry. Our simulations are 
among the few multidimensional simulations of Einstein equations in spherical 
coordinates. In the context of 3+1 formulations, some of the first simulations 
of black hole formation used spherical coordinates \cite{Stark85,Evans86}, 
however the formulations used in those works were unstable leading to 
exponentially growing constraint violations. Although some work has been done 
to reformulate the BSSN equations in order to ease its evolution in spherical 
coordinates \cite{Brown05, Alcubierre05, Alcubierre11}, these reformulations 
have been only tested for 1D numerical problems. The use of multi-block methods
in BSSN simulations \cite{Tiglio05, Pollney11} allows to make use of spherical 
wavezone grids to take advantage of topologically adapted grids, but they keep 
a cartesian grid in the central region.
On the other hand, spherical 
coordinates are widely used in the null formulations (see \cite{WinicourLR}), 
mostly in the context of Cauchy-characteristic matching and extraction (e.g. 
\cite{Reisswig09, Babiuc11, Reisswig11}), although stand-alone characteristic 
formulations have still very few numerical applications (e.g. \cite{Gomez98, 
Siebel02}).

We think that the reason for the success of our simulations in spherical 
coordinates is twofold. First, we use an implicit-explicit algorithm to solve 
the system of hyperbolic equations, whereas we solve implicitly the terms in 
the equations leading to instabilities. Second, only two degrees of freedom of 
the system, the GWs, are evolved by means of hyperbolic equations, while the 
rest are the result of the computation of elliptic equations. This main feature
of FCF is possibly crucial to provide the extra stability to the numerical 
algorithm. We are not sure, whether both requirements are indeed necessary to 
perform stable simulations in spherical coordinates, or whether the 
implicit-explicit scheme gives rise to the stability of the numerical 
algorithm. It would be interesting to explore the behavior of purely hyperbolic
formulations with our implicit-explicit algorithm in spherical coordinates. 

%%%%%%%%%%%%%%%%%%%%%%%%%%%%%%%%%%%%%%

\acknowledgments

We would like to thank the referees for their suggestions.
I. C.-C. acknowledges support from the Alexander von Humboldt Foundation. This 
work was also supported by the Collaborative Research Center on Gravitational 
Wave Astronomy of the Deutsche Forschungsgesellschaft (DFG SFB/Transregio 7),
by the grants AYA2010-21097-C03-01 of the Spanish MICINN and Prometeo 2009-103
of the Generalitat Valenciana. We would like to 
thank M.A. Aloy, J.A. Font, P. Montero, E. M\"uller and J. L. Jaramillo for 
their useful comments and discussion. 

%%%%%%%%%%%%%%%%%%%%%%%%%%%%%%%%%%%%%%

\appendix
\section{Runge-Kutta schemes and partially implicit evaluation of the $S_{w2}$ 
term}
\label{a:rk}

In this appendix we describe with more details the numerical method used in the
evolution of the variables $\bf W$, which has a partially implicit evaluation 
of the $S_{w2}$ term in the evolution of the $w^{ij}_k$ tensor. They are based 
in the explicit Runge-Kutta schemes of second, third and fourth-order. We show 
the procedure with the second and third-order ones.

The optimal second and third-order Runge-Kutta schemes~\cite{ShuOsher} of a 
general evolution equation in time $t$ for the variable $u$ of the form 
$u_t = L (u)$ are respectively
\bea
	u^{(1)} &=& u^n + \Delta t \, L(u^n), \nonumber \\
	u^{n+1} &=& \frac{1}{2} \left(u^n + u^{(1)} + \Delta t \, L (u^{(1)}) \right),
\eea
and
\bea
	u^{(1)} &=& u^n + \Delta t \, L(u^n), \nonumber \\
	u^{(2)} &=& \frac{3}{4} u^n + \frac{1}{4} u^{(1)} + \frac{1}{4} \Delta t \, L (u^{(1)}), \nonumber \\
	u^{n+1} &=& \frac{1}{3} u^n + \frac{2}{3} u^{(2)} + \frac{2}{3} \Delta t \, L (u^{(2)}),
\eea
where $u^n=u(t^n)$ and $u^{n+1}$ is the numerical approximation for the value 
$u(t^n+ \Delta t)$.

The corresponding methods used in order to evolve $\bf W$ based on the previous
Runge-Kutta schemes are:

\begin{enumerate}
\item Based on the second-order Runge-Kutta scheme:
\bea
	h^{ij(1)} &=& h^{ij(n)} + \Delta t \, S_h (h^{ij (n)}, \hat{A}^{ij (n)}, w^{ij (n)}_k, {\bf V}^{(n)}), \nonumber \\
	\hat{A}^{ij(1)} &=& \hat{A}^{ij(n)} + \Delta t \, S_{\hat{A}} 
(h^{ij (n)}, \hat{A}^{ij (n)}, w^{ij (n)}_k, {\bf U}^{(n)}, {\bf V}^{(n)}), \nonumber \\
	w^{ij(1)}_k &=& w^{ij(n)}_k + \Delta t \, S_{w1} (h^{ij(1)}, \hat{A}^{ij(1)}, {\bf V}^{(n)}) \nonumber \\
	&+& \Delta t \, S_{w2} (w^{ij(n)}_k, {\bf V}^{(n)}), \nonumber \\
	h^{ij(n+1)} &=& \frac{1}{2} h^{ij(n)} + \frac{1}{2} h^{ij(1)} + \frac{1}{2} \Delta t \, S_h (h^{ij(1)}, \hat{A}^{ij(1)}, w^{ij(1)}_k, {\bf V}^{(n)}), \nonumber \\
	\hat{A}^{ij(n+1)} &=& \frac{1}{2} \hat{A}^{ij(n)} + \frac{1}{2} \hat{A}^{ij(1)} \nonumber \\
	&+& \frac{1}{2} \Delta t \, S_{\hat{A}} 
(h^{ij (1)}, \hat{A}^{ij (1)}, w^{ij (1)}_k, {\bf U}^{(n)}, {\bf V}^{(n)}), \nonumber \\
	w^{ij(n+1)}_k &=& \frac{1}{2} w^{ij(n)}_k + \frac{1}{2} w^{ij(1)}_k + \frac{1}{2} \Delta t \, S_{w1} (h^{ij(n+1)}, \hat{A}^{ij(n+1)}, {\bf V}^{(n)}) \nonumber \\
	&+& \frac{1}{2} \Delta t \, S_{w2} (w^{ij(1)}_k, {\bf V}^{(n)}).
\eea
\vspace{2.cm}
\item  Based on the third-order Runge-Kutta scheme:
\bea
	h^{ij(1)} &=& h^{ij(n)} + \Delta t \, S_h (h^{ij (n)}, \hat{A}^{ij (n)}, w^{ij (n)}_k, {\bf V}^{(n)}), \nonumber \\
	\hat{A}^{ij(1)} &=& \hat{A}^{ij(n)} + \Delta t \, S_{\hat{A}} 
(h^{ij (n)}, \hat{A}^{ij (n)}, w^{ij (n)}_k, {\bf U}^{(n)}, {\bf V}^{(n)}), \nonumber \\
	w^{ij(1)}_k &=& w^{ij(n)}_k + \Delta t \, S_{w1} (h^{ij(1)}, \hat{A}^{ij(1)}, {\bf V}^{(n)}) \nonumber \\
	&+& \Delta t \, S_{w2} (w^{ij(n)}_k, {\bf V}^{(n)}), \nonumber \\
	h^{ij(2)} &=& \frac{3}{4} h^{ij(n)} + \frac{1}{4} h^{ij(1)} + \frac{1}{4} \Delta t \, S_h (h^{ij(1)}, \hat{A}^{ij(1)}, w^{ij(1)}_k, {\bf V}^{(n)}), \nonumber \\
	\hat{A}^{ij(2)} &=& \frac{3}{4} \hat{A}^{ij(n)} + \frac{1}{4} \hat{A}^{ij(1)} + \frac{1}{4} S_{\hat{A}} 
(h^{ij (1)}, \hat{A}^{ij (1)}, w^{ij (1)}_k, {\bf U}^{(n)}, {\bf V}^{(n)}), \nonumber \\
	w^{ij(2)}_k &=& \frac{3}{4} w^{ij(n)}_k + \frac{1}{4} w^{ij(1)}_k + \frac{1}{4} \Delta t \, S_{w1} (h^{ij(2)}, \hat{A}^{ij(2)}, {\bf V}^{(n)}) \nonumber \\
	&+& \frac{1}{4} \Delta t \, S_{w2} (w^{ij(1)}_k, {\bf V}^{(n)}) \nonumber \\
	h^{ij(n+1)} &=& \frac{1}{3} h^{ij(n)} + \frac{2}{3} h^{ij(2)} + \frac{2}{3} \Delta t \, S_h (h^{ij(2)}, \hat{A}^{ij(2)}, w^{ij(2)}_k, {\bf V}^{(n)}), \nonumber \\
	\hat{A}^{ij(n+1)} &=& \frac{1}{3} \hat{A}^{ij(n)} + \frac{2}{3} \hat{A}^{ij(2)} \nonumber \\
	&+& \frac{2}{3} \Delta t \, S_{\hat{A}} 
(h^{ij (2)}, \hat{A}^{ij (2)}, w^{ij (2)}_k, {\bf U}^{(n)}, {\bf V}^{(n)}), \nonumber \\
	w^{ij(n+1)}_k &=& \frac{1}{3} w^{ij(n)}_k + \frac{2}{3} w^{ij(2)}_k + \frac{2}{3} \Delta t \, S_{w1} (h^{ij(n+1)}, \hat{A}^{ij(n+1)}, {\bf V}^{(n)}) \nonumber \\
	&+& \frac{2}{3} \Delta t \, S_{w2} (w^{ij(2)}_k, {\bf V}^{(n)}).
\eea
\end{enumerate}
 
%%%%%%%%%%%%%%%%%%%%%%%%%%%%%%%%%%%%%%%%%

\end{document}